\documentclass[%
reprint,
 amsmath,amssymb,
 aps,
]{revtex4-2}

\usepackage{graphicx}
\usepackage{dcolumn}
\usepackage{bm}

\def\etmiss{\slashed{E}_T}

\usepackage{slashed}
\usepackage{mathtools}
\usepackage{graphicx}
\usepackage{braket}
\usepackage{caption}
\usepackage{subcaption}
\newcommand{\beq}{\begin{equation}}
\newcommand{\eeq}{\end{equation}}
\newcommand{\bea}{\begin{eqnarray}}
\newcommand{\eea}{\end{eqnarray}}

\def\ba{\begin{array}}
\def\ea{\end{array}}
\def\l{\lambda}

\def\mueff{\mu_\mathrm{eff}}

\def\sQ3{\widetilde{Q}_3}
\def\sU3{\widetilde{U}_3}
\def\sD3{\widetilde{D}_3}
\def\bino{\widetilde{B}}
\def\wino{\widetilde{W}}

\def\higgsinod{\widetilde{H}^0_d}
\def\higgsinou{\widetilde{H}^0_u}
\def\singlino{\widetilde{S}}
\def\vu{v_u}
\def\vd{v_d}
\def\vs{v_{_S}}
\newcommand{\fbinv}{\text{fb}$^{-1}$}

\begin{document}


\title{Intrinsic Geometry of Collider Observations and Forman Ricci Curvature}

\author{Jyotiranjan Beuria}%
 \email{jbeuria@iitmandi.ac.in}
\affiliation{Indian Institute of Technology Mandi, Himachal Pradesh, India}
\affiliation{IKS Research Center, ISS Delhi, Delhi, India}

\date{\today}

\begin{abstract}
We study the global and local topological properties of multi-lepton patterns reconstructed at the detectors. We investigate the sensitivity of Forman Ricci curvature distributions and persistent homology features to kinematic cuts, integrated luminosity, and scales of maximum filtration. We find that these topological properties are efficient enough in discriminating the BSM scenarios from the SM background, particularly when the BSM scenarios possess a massive invisible particle. We also find that the topological properties exhibit a scaling behaviour with integrated luminosity. This exploratory study suggests that the topological features can potentially supplement the traditional cut-and-count analyses in search of new physics.

\end{abstract}

\maketitle


\section{\label{sec:intro}Introduction}

The Standard Model (SM) of particle physics has proven highly effective in explaining the behaviour of elementary particles. Identifying the Higgs boson, a neutral scalar particle, marked a significant achievement for the SM \citep{aad2012observation,chatrchyan2012observation}. Despite its success, the SM needs to address several crucial unanswered questions. Consequently, searching for models beyond the SM  has prompted various upgrades to the Large Hadron Collider (LHC) and the emergence of other future colliders. 

The exploration of the Beyond the Standard Model (BSM) phenomenology involves constraining the parameter space of new physics models. This is achieved through collider simulations at the parton, hadron, and detector levels, incorporating signal and background analyses. Usually, it is accepted that collider events are independent and, to a great extent, identically distributed as well. However, the reconstructed objects at the detector might preferentially populate some portion of the phase space depending on the available phase space for the decaying heavy BSM particles. Thus, physicists have introduced a range of kinematic variables \cite{franceschini2023kinematic} for this purpose, relying on kinematic cuts applied on an event-by-event basis. The traditional cut-and-count method has been immensely successful in excluding several BSM scenarios. However, the intrinsic geometric structure of the phase space of the reconstructed particles featuring higher-order topological properties might add to ongoing searches at the colliders. These, coupled with traditional approaches, can potentially draw stronger exclusion limits even at low values of integrated luminosity.

In recent years, several efforts have been made to analyze the global patterns of the collider observations using techniques such as Voronoi and Delaunay tessellations \cite{Debnath:2015wra, Debnath:2016mwb, Matchev:2020vhr}, network distance metrics \cite{Mullin:2019mmh}, and machine learning based classifications \cite{guest2018deep, Mullin:2019mmh,qasim2019learning,abdughani2019probing,du2020identifying,flesher2021parameter, Chang:2020rtc,  Nachman:2021yvi, knipfer2023deep}. Recently, Topological Data Analysis (TDA) \cite{taylor2015topological, topaz2015topological, lloyd2016quantum, gidea2018topological, saggar2018towards, sizemore2019importance, murugan2019introduction, cole2019topological, chazal2021introduction,beuria2023persistent,gupta2024characterizing} is being increasingly used in the broader field of data science for examining the intrinsic global properties of the system. However, its utility in discriminating the BSM signals from the SM background is not well understood.

TDA is a mathematical and computational approach that applies tools from algebraic topology to analyze the intrinsic topology of complex datasets. This allows for a more robust understanding of the underlying space and relationships within the dataset. As an alternative to traditional machine learning approaches, TDA is particularly effective in studying the global properties and connectivity of data points. It introduces concepts such as persistent homology, which helps identify significant topological features that persist across different scales. By representing data as a topological space, TDA enables the extraction of valuable information about clusters, voids, and other topological structures that may not be apparent through other analytical methods. In practical applications, TDA has been useful in biology, neuroscience, machine learning, and materials science. It offers a robust framework for uncovering hidden patterns and structures in diverse datasets, providing insights that traditional methods might overlook. 

Since the traditional TDA technique uses unweighted simplices, a great deal of information encoded in the global geometry of the dataset is likely to be unrecoverable. By assigning weights to simplices, TDA can distinguish between strong and weak connections, emphasizing essential features and filtering out noise. This enhances the ability to extract meaningful information about the global and local structures of the dataset. 

We also explore another important topological property, Ricci curvature of the simplicial complexes. Ricci curvature, a fundamental concept in differential geometry and physics, characterizes local manifold properties, such as the volume of distance balls and geodesic divergence. In general relativity, the Einstein field equations link space-time geometry to matter distribution using the Ricci curvature tensor. Ricci flow, integral to Perelman's proof of the Poincaré conjecture \cite{perelman2003ricci}, further underscores its significance. Discrete Ricci curvature forms \cite{samal2018comparative}, namely Ollivier Ricci curvature (ORC) \cite{ollivier2007ricci,ollivier2009ricci} and Forman Ricci curvature (FRC) \cite{forman2003bochner,sreejith2016forman}, extend the classical concept to networks and simplicial complexes. ORC, based on Wasserstein distance, captures clustering in network structures, while FRC, derived combinatorially, reveals geodesic dispersal and topological information. Despite potential dissimilarities, ORC and FRC exhibit high correlations in complex networks. ORC proves effective for probabilistic analyses, while FRC excels in understanding combinatorial network properties. In this study, we explore the FRC distributions.

Particle colliders like LHC collect massive amounts of data featuring various production and decay cascades of elementary particles. Physicists have studied the phenomenology of several SM extensions, viz., Supersymmetry \cite{csaki1996minimal, martin1998supersymmetry, Baer:2006rs, ellwanger2010next} and Minimal Universal Extra Dimension \cite{cembranos2007exotic, datta2010minimal, beuria2018lhc}, Two Higgs Doublet models \cite{davidson2005basis,branco2012theory}, etc. to name a few. In a recent exploratory work \cite{beuria2023persistent}, we have studied the topological properties of the reconstructed objects using unweighted persistent homology for a minimal extension of the SM by a real singlet scalar \cite{ham2005electroweak, barger2008cern, guo2010real}.

In order to illustrate the usefulness of topological patterns in distinguishing the BSM scenarios and the SM background, we choose leptons from the resonant production of light neutralino and chargino in the Next-to-Minimal Supersymmetric Standard Model (NMSSM). One of the primary focuses of this study is to demonstrate how the topological information behaves across different kinematic cuts, scales of maximum filtration, and integrated luminosity values. We discuss three topological features: Forman Ricci curvature (FRC) distributions, persistent entropy, and persistent amplitude for three benchmark scenarios in the NMSSM. The discussion is quite generic and can be extended to any collider observations. FRC distributions capture the topological properties at a particular scale of filtration, and persistent homology studies it on multiple scales.

The organization of the paper is as follows. In section \ref{sec:wph}, we present some mathematical preliminaries of topological data analysis. We give the basic framework of analysis in section \ref{sec:framework}. We discuss the variation of the topological features and signal-background classification accuracy with kinematic cut selection and integrated luminosity in section \ref{sec:features}. We conclude the discussion in section \ref{sec:conclusion}.

\section{Topological Data Analysis (TDA)}
\label{sec:wph}
\begin{figure}[t]
\centering 
\includegraphics[width=0.45\textwidth, angle=0]{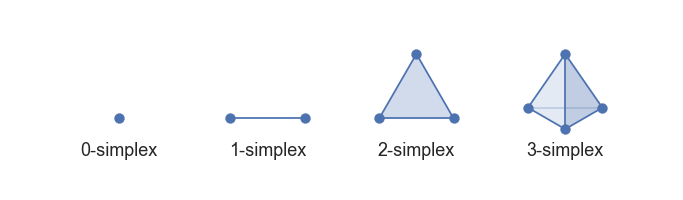}	
\caption{Simplices (plural of simplex) are the combinatorial building blocks of a simplicial complex. For illustration, 0-,1-,2-, and 3-simplex are shown from left to right.} 
\label{fig:simplex}%
\end{figure}
One of the prominent tools used for topological data analysis is persistent homology. The foundational geometric structure for studying persistent homology is the simplicial complex. Simplicial complexes provide a method for constructing topological spaces using fundamental combinatorial building blocks called simplices. It simplifies the treatment of the continuous geometry of topological spaces. Instead, it involves the more manageable tasks of combinatorics and counting. These elementary building blocks, known as simplices and illustrated in Fig. \ref{fig:simplex}, are formed by taking the convex hull of independent points. A $k$-dimensional simplex is generated by this process, involving $k+1$ points. For instance, a 0-simplex is a point, a 1-simplex is a line segment, a 2-simplex is a filled triangle, and a 3-simplex is a filled tetrahedron. This construction can be extended to higher-dimensional polytopes. In an $n$-dimensional simplex, the simplices with dimensions $k<n$ constitute its faces. Consequently, for a 2-simplex (triangle), its edges (1-simplex) serve as the faces. Throughout this work, we have used the Vietoris-Rips complex to study the topology. It constructs simplicial complexes from point cloud data, connecting points within a specified distance threshold, known as the filtration parameter or filtration scale.


\subsection{Persistent Homology Features}
The $k$-th homology group, denoted as $H_k$, is the quotient group, representing cycles modulo boundaries. Mathematically, it is expressed as:
\begin{equation}
    H_k = \frac{Z_k}{B_k} = \frac{\text{ker}(\partial_k)}{\text{im}(\partial_{k+1})} 
\label{eq:hk}
\end{equation}
Here, $H_k(K)$ is the quotient vector space whose generators are given by $k$-cycles that are not boundaries of any $(k+1)$-simplices. The rank of $H_k(K)$ is referred to as the $k$-th Betti number, denoted as $\beta_k(K)$. The Betti number $\beta_k(K)$ signifies the number of $k$-dimensional holes in the simplicial complex $K$ that are not boundaries of any $(k+1)$-simplices. For instance, $\beta_0(K)$ represents the number of connected components in $K$. It is important to note that Betti numbers $\beta_k$ define the Euler characteristic ($\chi$), a topological invariant of the simplicial complex, given by:
\begin{equation}
\chi = \sum_{k=0}^{n} (-1)^k \beta_k 
\label{eq:betti}
\end{equation}

Another important measure is the entropy of the points clustered in the so-called persistence diagram, also called persistence entropy. The persistence diagram shows the appearance ($birth$) and disappearance ($death$) of $k-$th hole as the filtration parameter changes. Let $D=\{ (b_i,d_i) \}$ be the set of all $birth$-$death$ pairs associated with $k-$th order homology group in persistence diagram with $d_i < \infty$. The $k-$th order persistence entropy is given by
\begin{equation}
    S_k^{pe}=S(D_k)=-\sum_{i}p_i \log(p_i), 
\end{equation} 
where $p_i=\frac{d_i-b_i}{L_D}$ and $L_D=\sum_{i}{(d_i-b_i)}$.

We also make use of a topological feature, persistent amplitude defined on $D$, the set of persistent ($birth$,$death$) pairs as a function 
\begin{align*}
    A :\; D \rightarrow R,
\end{align*}
for which there exists a vectorization 
\begin{align*}
    \Phi :\; D \rightarrow V,
\end{align*}
with $V$, a normed space such that 
\begin{align*}
    A(x) &= \|\Phi(x)\|
\end{align*}
for all $x \in D$. We choose the Wasserstein metric to compute distances between persistent diagrams and amplitude. The choice of the Wasserstein metric is primarily motivated by its central stability properties \cite{cohen2010lipschitz,skraba2020wasserstein,songdechakraiwut2023wasserstein}.

\subsection{Forman Ricci Curvature}

Let $\alpha$ and $\bar{\alpha}$ be $k$-dimensional simplices  in the simplicial complex $K$. If there exists a simplex $\beta$ in $K$ such that $\beta > \alpha$ and  $\beta > \bar{\alpha}$, $\alpha$ and $\bar{\alpha}$ have a common co-face $\beta$ and they are termed as upper adjacent. Similarly, $\alpha$ and $\bar{\alpha}$ are said to be lower adjacent if they share a common face $\gamma$ (a $(k-1)$-simplex), that is, $\gamma < \alpha$ and  $\gamma < \bar{\alpha}$. If $\alpha$ and $\bar{\alpha}$ are either lower or upper adjacent, but not both, they are said to be parallel. Then, Forman Ricci Curvature (FRC) is given by \cite{forman2003bochner}
\begin{eqnarray}
R_k(\alpha)= N(\textrm{Upper adjacent simplices}) \nonumber \\ 
            + N (\textrm{Lower adjacent simplices}) \nonumber  \\
            -N(\textrm{Parallel simplices}) 
\end{eqnarray}

In case of weighted simplicial complexes with weights $w$, $R_k(\alpha)$ is given by

\begin{equation}
    \begin{aligned}
R_k(\alpha) =& w_{\alpha} \left[ \sum_{\beta > \alpha} \frac{w_{\beta}}{w_{\alpha}} + \sum_{\gamma < \alpha} \frac{w_{\gamma}}{w_{\alpha}} \right ] \\ 
-& w_{\alpha} \sum_{\bar{\alpha} \neq \alpha} \left[ \sum_{\beta > \alpha,\bar{\alpha}}  \frac{\sqrt{w_{\alpha}w_{\bar\alpha}}} {w_{\beta}} - \sum_{\gamma < \alpha,\bar{\alpha}}  \frac {w_{\gamma}} {\sqrt{w_{\alpha}w_{\bar\alpha}}} \right] 
\end{aligned}
\end{equation}

For an edge, Forman Ricci curvature reduces to
\begin{equation}
    \begin{aligned}
R_1(\alpha) =& w_{\alpha} \left(  \sum_{\gamma < \alpha} \frac{w_{\gamma}}{w_{\alpha}}  -  \sum_{\bar{\alpha} \neq \alpha}  \sum_{\gamma < \alpha,\bar{\alpha}}  \frac {w_{\gamma}} {\sqrt{w_{\alpha}w_{\bar\alpha}}} \right)
\end{aligned}
\end{equation}

\begin{table*}[!ht]
\centering
\begin{tabular}{|c|c|c|c|c|c|c|c|c|c|c|c|c|} 
 \hline 
   & $\mueff$ & $\l$ & $\kappa$ & $A_\lambda$ & $A_\kappa$ & $m_{h_{1,2}}$ & $m_{\chi_{1,2,3}^0}$ & $m_{\chi^\pm}$ & $BR_{\chi_2^0\rightarrow \chi_1^0\:Z}$ & $BR_{\chi_3^0\rightarrow \chi_1^0\:Z}$ & $BR_{\chi_3^0\rightarrow \chi_1^\pm\:W^\mp}$ & $BR_{\chi_1^\pm \rightarrow \chi_1^0\:W^\pm}$ \\ \hline
  
  BP1 & 180 & 0.6 & -0.24 & 1000 & 300 & 95.6, 124.0 & 83.5, 191.3, 262.2 & 184.5 & 0.99 & 0.30 & 0.0 & 1.0 \\ \hline

  BP2 & 150 & 0.65 & -0.30 & 860 & 310 & 94.1, 124.3 & 61.6, 163.8, 251.0 & 153.7 & 0.97 & 0.08 & 0.85 & 1.0 \\ \hline

  BP3 & 250 & 0.69 & -0.35 & 1250 & 250 & 125.4, 233.0 & 157.1, 259.2, 369.5 & 255.7 & 1.0 & 0.00 & 0.80 & 1.0 \\ \hline
    
\end{tabular}
\caption{BP1, BP2 and BP3 are three benchmark scenarios in the NMSSM having two different values of $\mueff$ leading to $3\ell$ signals from the resonant production of $\chi_{2,3}^0\chi_1^\pm$. Recent LHC searches have excluded BP1 and BP2.}
\label{tab:benchmarks}
\end{table*}

\section{Framework of Analysis}
\label{sec:framework}
\subsection{$Z_3$ symmetric Next-to-Minimal Supersymmetric Standard Model (NMSSM)}
We choose the Next-to-Minimal Supersymmetric Standard Model (NMSSM) to demonstrate the utility of topological information in BSM searches. The NMSSM framework incorporates an additional singlet superfield denoted as $\hat{S}$ alongside the standard MSSM superfields. In the widely studied $Z_3$-symmetric variant of the NMSSM, the linear and bilinear terms in $\hat{S}$ are dropped. Additionally, the $Z_3$ symmetry restricts the inclusion of explicit higgsino mass term ($\mu$-term) in the NMSSM superpotential. The $\mu$-term is generated after the singlet scalar gets vev ($\mueff=\l \vs$). The $Z_3$-symmetric superpotential of the NMSSM is defined as:
\beq
\mathcal{W}=\mathcal{W_{MSSM}}|_{\mu=0}+ \lambda \hat{S} \hat{H}_u.\hat{H}_d
  + {\kappa \over 3} \hat{S}^3
\label{eq:superpot}
\eeq
with
\beq
\mathcal{W_{MSSM}}|_{\mu=0}= y_d \hat{H}_d\cdot\hat{Q} \hat{D}_R^c 
                           + y_u \hat{Q} \cdot  \hat{H}_u\hat{U}_R^c 
                           + y_e \hat{H}_d \cdot \hat{L} \hat{E}_R^c, \quad
\eeq
where ${\cal{W_{MSSM}}}|_{\mu=0}$ denotes the MSSM superpotential with the exclusion of the $\mu$-term. The superfields $\hat{H}_{u}$ and $\hat{H}_{d}$ correspond to the doublet Higgs superfields, while $\hat{S}$ represents the gauge singlet superfield mentioned earlier. The superfields $\hat{Q}$, $\hat{U}_R$, and $\hat{D}_R$ refer to the $SU(2)$ quark-doublet, up-type $SU(2)$ singlet quark, and down-type $SU(2)$ singlet quark superfields, respectively. Additionally, $\hat{L}$ and $\hat{E}_R$ represent the $SU(2)$ doublet and singlet lepton superfields, respectively. The symbols $y_{f=d,u,e}$ denote the corresponding Yukawa couplings.

The $5 \times 5$ symmetric neutralino mass matrix, in the basis 
\{$\bino, \wino, \higgsinod, \higgsinou, \singlino$ \}, is given by
\cite{ellwanger2010next}
\begin{equation}
 \label{eq:mneut}
 {\cal M}_{\chi^0} =
\left( \begin{array}{ccccc}
M_1 & 0 & -\frac{g_1 \vd}{\sqrt{2}} & \frac{g_1 \vu}{\sqrt{2}} & 0 \\
& M_2 & \frac{g_2 \vd}{\sqrt{2}} & -\frac{g_2 \vu}{\sqrt{2}} & 0 \\
& & 0 & -\mueff & -\l \vu \\
& & & 0 & -\l \vd \\
& & & & 2 \kappa \vs
\end{array} \right) \;,
\end{equation}

and the chargino sector is given by

\begin{equation}
 \label{eq:mneut}
 {\cal M}_{\chi^\pm} =
\left( \begin{array}{cc}
M_2 & g_2 \vu \\
g_1 \vd & \mueff
\end{array} \right) \;,
\end{equation}
where $M_1$ and $M_2$ represent the soft SUSY-breaking masses corresponding to the $U(1)$ ($\bino$) and $SU(2)$ ($\wino$) gauginos, respectively. The parameters $g_1$ and $g_2$ denote the respective gauge couplings associated with these gauginos. It is noteworthy that there is no direct mixing observed among the gauginos ($\bino$ and $\wino$) and the singlino ($\singlino$). However, a slight mixing is indirectly introduced through the neutral higgsino sector ($\higgsinod, \, \higgsinou$). Conversely, direct mixing between the higgsinos and the singlino can occur via the off-diagonal terms of ${\cal M}_{\chi^0}$ that are proportionate to $\lambda$. Consequently, scenarios characterized by relatively small $\mueff$ lead to lighter neutralinos exhibiting a notable blend of singlino and higgsino components across significant regions of the NMSSM parameter space. 

We present three scenarios ($\mueff \leq 250$ GeV) in Table \ref{tab:benchmarks} such that the LSP is singlino-like and the NLSP is higgsino-like along with a higgsino-like chargino.  Such low $\mueff$ also features a light singlet Higgs boson and the SM Higgs boson. All sfermions soft-breaking mass parameters are kept at 3 TeV. The particle spectra are calculated using \texttt{NMSSMTools v6.0.2} \cite{das2012nmsdecay}.

BP1 and BP2 benchmark scenarios are excluded by recent LHC searches \cite{ATLAS:2022zwa}. We choose these to suggest possible validation of using topological information in BSM searches using existing data. Also, the light charginos and higgsinos feature a sizable cross-section to illustrate topological signatures for $3\ell$ signals across an extended range of integrated luminosity and kinematic cuts. However, the discussed framework is generic enough for any BSM scenarios.

\subsection{Collider Simulation}
We consider resonant production of $\chi_{2,3}^0\chi_1^\pm$ and subsequent leptonic decay via $Z$ and $W^\pm$ bosons. This leads to $3\ell +\etmiss$ signature at the collider. We consider three potential contributors to the background, namely, $t \bar{t} Z$,  $t \bar{t} W^\pm$ and $ZW^\pm$ via leptonic decay of   $Z$ and $W^\pm$ bosons. However, for all practical purposes, the leptonic decay of  $ZW^\pm$ is the most dominant background for the  $3\ell +\etmiss$ signature. We veto the b-tagged jets after detector-level simulation.

Event samples are generated at the lowest order (LO) in perturbation theory using \texttt{MadGraph5 aMC@NLO v3.5.1} \cite{Alwall:2014hca, Frederix:2018nkq} with the \texttt{nn23lo1} \cite{ball2017parton} parton distribution function at $\sqrt{s} = 13$ TeV. The generated parton-level events undergo showering with \texttt{Pythia v8.309} \cite{bierlich2022comprehensive}. To avoid double counting of events in the simulated samples, especially in the presence of extra hard partonic jets and the parton shower, the event generator utilizes the MLM matching technique with the variables \textit{xqcut} and \textit{qcut} set at appropriate values. The cross sections for all processes are estimated using an NLO K-factor of 1.2.

The \texttt{FastJet (v3.3.4)} \cite{Cacciari:2005hq, Cacciari:2011ma} package, integrated into \texttt{Delphes v3.5.0} \cite{deFavereau:2013fsa}, is employed for jet finding. The anti-$k_T$ jet algorithm is utilized with a cone size of 0.5, requiring a minimum $p_T^{jet}$ of 20 GeV and limiting the pseudorapidity to $|\eta_{jet}| < 2.5$.

Following the default parameter settings of \texttt{Delphes v3.5.0}, the reconstruction of leptons (electrons and muons) involves a minimum $p_T^{l}$ of 10 GeV and $|\eta_{jet}| < 2.5$. For electrons and muons, the track isolation requirement entails removing jets within an angular distance $\Delta R \leq 0.5$ from the lepton. To enhance the purity of electrons, it is required that the ratio of the total $p_T$ of stray tracks within the cones of their identification to their own $p_T$ is less than 0.12. Similarly, the corresponding ratio for muons is set at 0.25.

\begin{figure*}[!ht]
  \centering
  \subfloat[]{\includegraphics[width=0.45\textwidth]{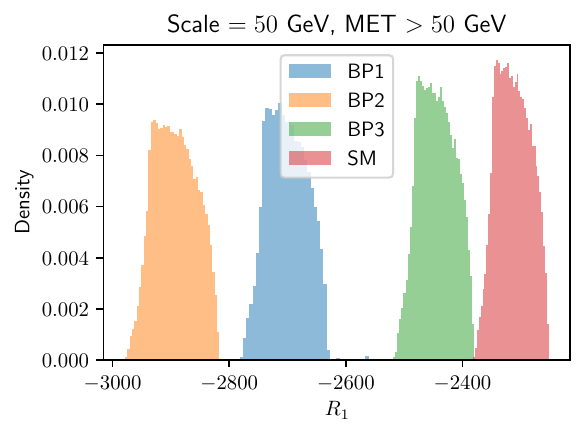}}
  \subfloat[]{\includegraphics[width=0.45\textwidth]{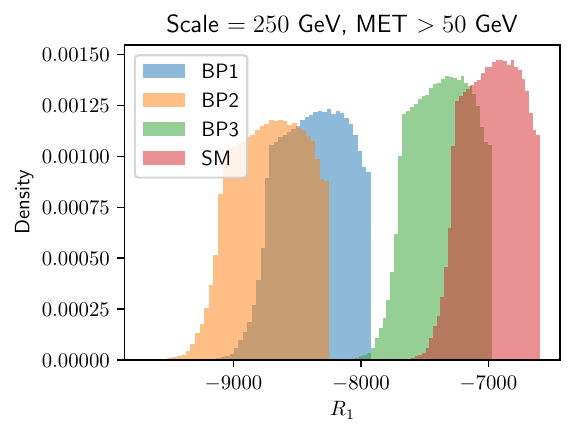}}

  \vskip 0.1in
  \subfloat[]{\includegraphics[width=0.45\textwidth]{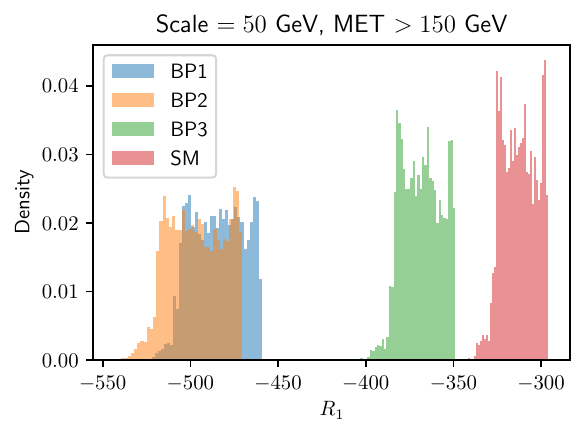}}
  \subfloat[]{\includegraphics[width=0.45\textwidth]{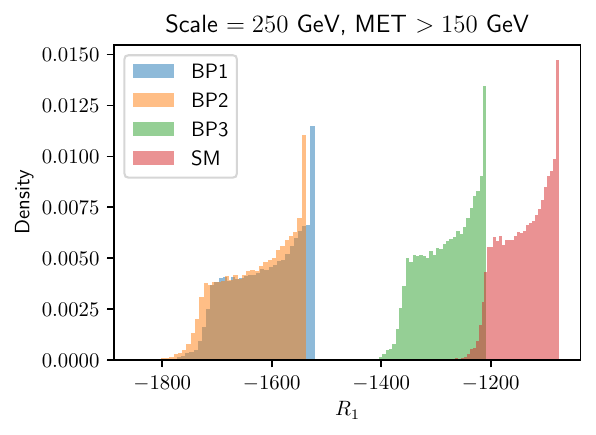}}
  \caption{$R_1$ distributions associated with different combinations of scales (50 GeV and 250 GeV) and missing transverse momentum (MET) cuts (50 GeV and 150 GeV). Integrated luminosity is fixed at 150 fb$^{-1}$.}
  \label{fig:hist_ricci}
\end{figure*}

We choose two values of MET, i.e., MET $> 50~\mathrm{GeV}$  and MET $> 150~\mathrm{GeV}$ to demonstrate the impact of kinematic cuts. We require at least three hard leptons ($\geq 20$ GeV) and veto any b-tagged jets.
We require $p_T^\ell \geq 50~\mathrm{GeV}$ for the first $p_T$ ordered lepton. The kinematic cuts are implemented using \texttt{MLAnalysis} \cite{guo2024mlanalysis} library.  Table \ref{tab:xsection} presents the cross-section and cut-selection details at $\sqrt{s} = 13$ TeV and 150 \fbinv of integrated luminosity. The significance level is calculated using $\frac{S}{\sqrt{S+B}}$, where S and B represent the number of the signal events and the SM background events surviving after the kinematic cuts have been applied.

\begin{table}[]
\centering
\begin{tabular}{|l|l|l|l|l|}
\hline
    & Cross Section & $N_{cut}$ & Significance  & Status\\ \hline
SM  & 435.1  fb    & 576 (96)      &         &      \\ \hline
BP1 & 7.90  fb      & 116 (39)     & 4.41 (3.36) & Excluded      \\ \hline
BP2 & 13.53 fb     & 142 (41)    & 5.30 (3.50)  &  Excluded     \\ \hline
BP3 & 2.41  fb      & 33  (12)      & 1.34 (1.15) & Not Excluded        \\ \hline
\end{tabular}
\caption{Production cross section along the leptonic channel and cut selection ($N_{cut}$) of the SM background and the benchmark scenarios at 13 TeV and 150 \fbinv for MET $> 50~\mathrm{GeV}$ ($> 150~\mathrm{GeV}$).}
\label{tab:xsection}
\end{table}

\subsection{Point cloud data from collider observations}
After the fast detector level simulation, we collect the reconstructed leptons event-wise. The number of events is also normalized to the integrated luminosity multiplied by the effective cross-section of the considered process. Integrated luminosity values are varied between 50 \fbinv to 1000 \fbinv to understand the evolution of topological properties and their discriminatory power in search of new physics.

The point cloud comprises all leptons passing the basic kinematic cuts mentioned above. The three momenta of the reconstructed leptons serve as the coordinate for points in the point cloud. Thus, a single collider event passing the above-mentioned cuts contributes three vertices to the Vietoris-Rips complex because of our leptonic requirements. For simplicity, we use cartesian basis for the leptonic momentum.

In the study of persistent homology of the Vietoris-Rips complex, the filtration parameter is the scale. Ricci curvature captures the local behaviour of the network connectivity at a particular scale. However, persistent homology features such as the Betti curve, persistent entropy, and persistent amplitude are essentially multi-scale features. They represent the global behaviour of the system across different scales. Therefore, we employ two slightly different strategies while studying Ricci curvature and persistent features, as discussed below. 

It is to be noted that we do not need the coordinates of vertices explicitly to find the Forman Ricci curvature distributions. All we need to know is whether an edge exists between two vertices and the weights assigned to them. We choose an edge (i,j) to exist only if $\mathrm{max}\{p_T^{\ell_i},p_T^{\ell_j}\} \leq \mathrm{scale}$.  We choose a weighting system motivated by the collider observations. For simplicity, the vertices are assigned unit weights. The edges are assigned a weight that depends on the ratio of $p_T$ of the leptons forming the edge. It is expressed as follows.

\begin{align}
 \label{eq:weights}
 w_i &= 1 \nonumber \\
 w_{ij} &= \sqrt{\frac{p_T^{\ell_i} + p_T^{\ell_j}}{\mathrm{max}\{p_T^{\ell_i},p_T^{\ell_j}\}}}
\end{align}
However, we have used unweighted simplicial complexes to study the persistent homology of collider observations. We use our custom code to compute Ricci curvature distributions and \texttt{Giotto-ai v0.6.0} library \cite{giotto-tda} for persistent homology features.

\begin{figure*}[!ht]
  \centering
  \subfloat[]{\includegraphics[width=0.45\textwidth]{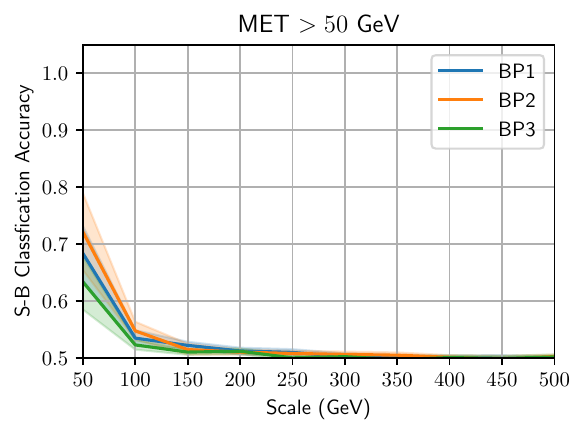}}
  \subfloat[]{\includegraphics[width=0.45\textwidth]{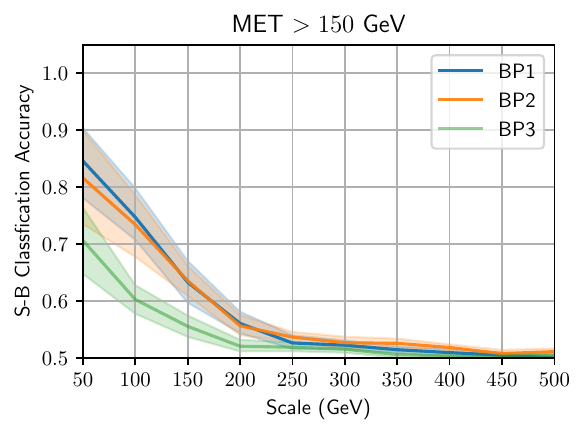}}
  \caption{Signal-Background classification accuracy across different 
  scales at 150 \fbinv using $R_1$ for two values of missing transverse momentum (MET) are shown. The shaded regions correspond to the 95\% CL.}
  \label{fig:acc_R1_IL150_scales}
\end{figure*}

\subsection{Machine learning and feature extraction}

Given a Vietoris-Rips complex at a particular scale, we obtain the Ricci curvature distributions for the edges and use this as an input for machine learning classification of the topological features. Thus, the input feature is a one-dimensional vector whose length is the same as the number of edges in the complex. Throughout this study, we compare the topological features of the background with the combined events of background and signal. Thus, the sample is essentially skewed. Thus, we perform random under-sampling so that the training of the classifier is not biased. Without this important operation, the binary classification accuracy almost always becomes 100\%. We also normalize the input data using standard scalar normalization or $Z-$score normalization. 

We use Support Vector Machine (SVM) with the radial basis function (RBF) kernel. SVM is a powerful supervised learning algorithm used for both classification and regression tasks. The regularization parameter, also known as the penalty parameter, $C$, is fixed at 1.0 throughout this study. It controls the trade-off between maximizing the margin and minimizing the classification errors.  We also perform 10-fold cross-validations to reduce any possible statistical fluctuation in ML classifications. 

In the latter part of this work, we use Betti number distributions across filtration scales up to 50 GeV or 250 GeV for ML classification with SVM. The Betti number distributions are also popularly called Betti curves. We determine the Betti numbers up to the second homology dimensions. Thus, the input data for SVM is three-dimensional in this case. The input parameters used in SVM are the same as before.

\section{Results and Discussion}
\label{sec:features}
This study aims to explore the dependence of topological properties of the leptonic distributions on kinematic cut selection and integrated luminosity. Many BSM frameworks, like NMSSM, feature a massive LSP. Thus, we expect the leptonic momentum distributions for the BSM to differ from the SM background. Since the mass of the invisible particle affects the missing transverse momentum distributions, the scale of maximum filtration is likely to be correlated with it. 

We also explore the signal-background discriminatory potential of topological properties in relation to kinematic cuts on MET, integrated luminosity values, and the maximum filtration scale. We find local topological features using Forman Ricci curvature and global topological properties through persistent homology. 

As the subsequent analysis will reveal, these intrinsic geometric properties of collider observations are efficient enough to discriminate new physics scenarios from the SM background. We have also explicitly checked that the results reported below are not very much sensitive to experimental systematic errors (at 5\% level) as these fail to alter the topological structure of the dataset appreciably.

\subsection{Forman Ricci Curvatue}

\begin{figure*}[!ht]
  \centering
  \subfloat[]{\includegraphics[width=0.45\textwidth]{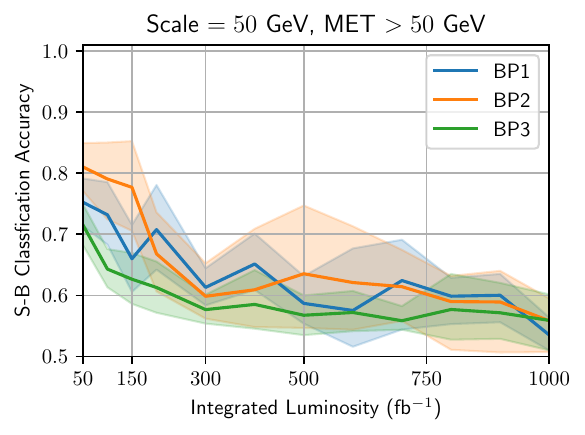}}
  \subfloat[]{\includegraphics[width=0.45\textwidth]{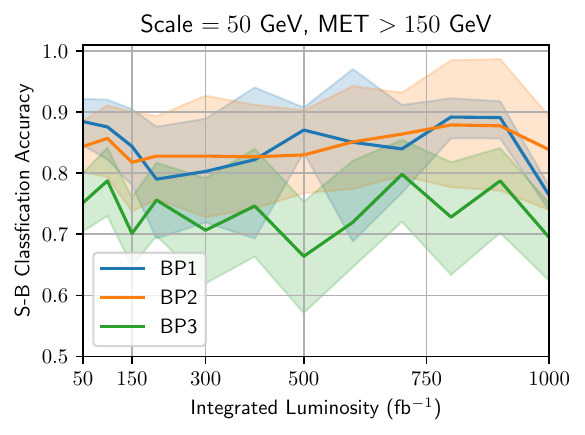}}
  \caption{Classification accuracies across different 
  integrated luminosity using $R_1$ for two values of missing transverse momentum (MET) are shown. The scale of filtration is fixed at 50 GeV. The shaded regions correspond to the 95\% CL.}
  \label{fig:acc_R1_IL150_lums}
\end{figure*}

In Fig. \ref{fig:hist_ricci}, we present the Forman Ricci curvature $R_1$ distribution of the three benchmark scenarios and the SM background for two scales (50 GeV and 250 GeV) and for two cuts on missing transverse momentum (MET $> 50$ GeV and MET $> 150$ GeV) at 150 \fbinv integrated luminosity. We observe that the Ricci curvature distribution for BP2 occupies most negative values, and the distribution for BP3 is very close to the SM case. This is primarily because of the cross-section distributions. It is to be reminded that we are considering a combined sample of signal and background for the benchmark scenarios. 

Comparing Fig. \ref{fig:hist_ricci}(a) and Fig. \ref{fig:hist_ricci}(c), we note that with increasing MET, the Ricci curvature distributions for BP1 and BP2 become less distinct from each other. MET $> 150$ GeV eliminates most of the SM background. Thus, the edge connectivity is reduced significantly which results in less negative $R_1$ compared to the MET $> 50$ GeV case.

However, increasing the scale of filtration to 250 GeV increases the edge connectivity for a node. Thus, the SM distribution starts overlapping with the BSM ones. This is shown in Fig. \ref{fig:hist_ricci}(b) and Fig. \ref{fig:hist_ricci}(d). The scale at 250 GeV admits more edges compared to 50 GeV. Thus, for both MET $> 50$ GeV and MET $> 150$ GeV, the $R_1$ is more negative than their counterparts at the 50 GeV scale. As before,  MET $> 150$ GeV reduces more SM events than BSM scenarios. Thus, Fig. \ref{fig:hist_ricci}(c) and Fig. \ref{fig:hist_ricci}(d) demonstrate qualitatively similar behaviour.

In Fig. \ref{fig:acc_R1_IL150_scales}, we present signal and background classification across different scales for maximum filtration using the SVM framework, as discussed previously. Fig. \ref{fig:acc_R1_IL150_scales}(a) corresponds to the MET $> 50$ GeV scenario, and Fig. \ref{fig:acc_R1_IL150_scales}(b) corresponds to the MET $> 150$ GeV scenario. We notice that the classification accuracy remains remarkable for both at lower scales. However, with the increasing scale of maximum filtration, accuracy drops. In Fig. \ref{fig:acc_R1_IL150_scales}(b), the accuracy remains significant until a larger scale. We can attribute this to the better signal-background separation at higher MET. Thus, the scale of maximum filtration is inevitably connected with the MET and, thereby, the mass of the LSP.

\begin{figure*}[!ht]
  \centering
  \subfloat[]{\includegraphics[width=0.3\textwidth]{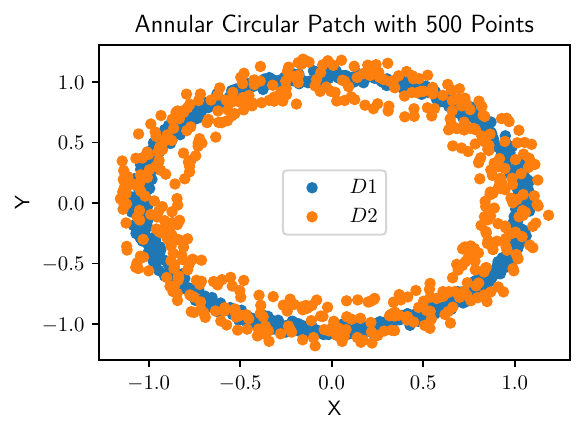}}
  \subfloat[]{\includegraphics[width=0.3\textwidth]{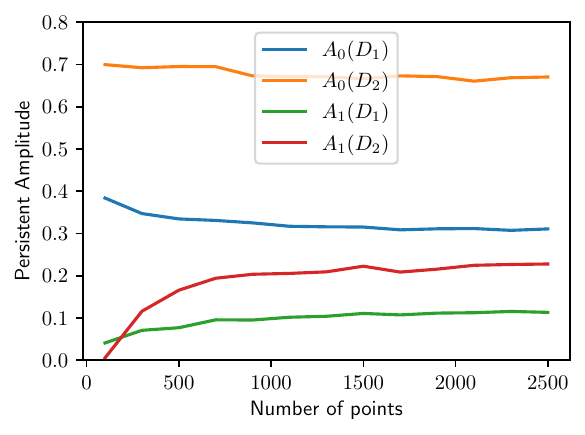}}
  \subfloat[]{\includegraphics[width=0.3\textwidth]{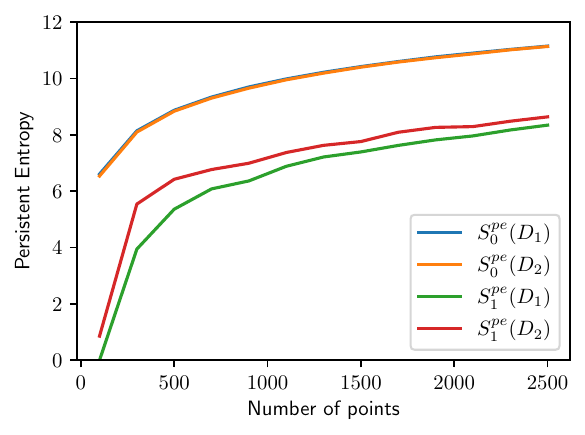}}

  \vskip 0.1in
  \subfloat[]{\includegraphics[width=0.3\textwidth]{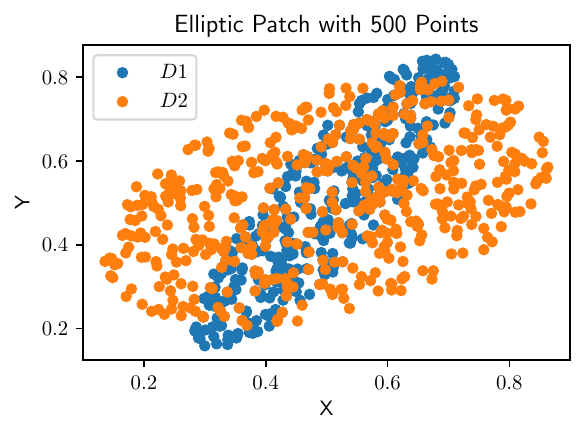}}
  \subfloat[]{\includegraphics[width=0.3\textwidth]{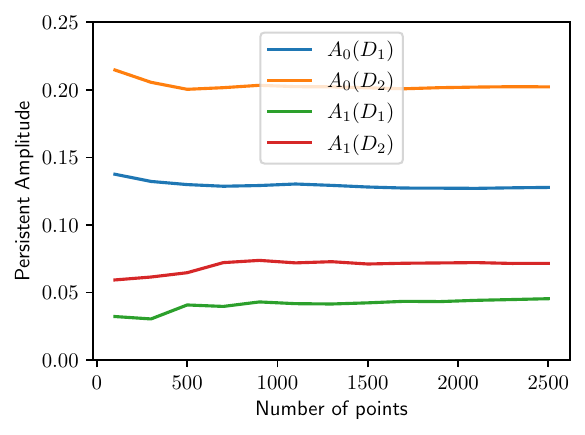}}
  \subfloat[]{\includegraphics[width=0.3\textwidth]{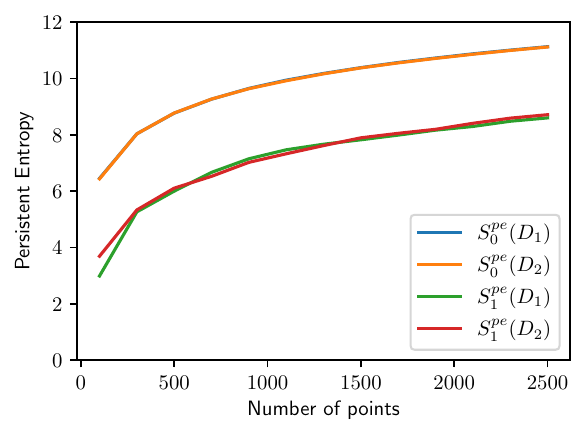}}

  \caption{The first row presents two distributions ($D_1$ and $D_2$) sampled from the annular regions of different radii. The second row presents points sampled from elliptical patches of different widths. The number of points in $D_1$ and $D_2$ are always equal.}
  \label{fig:demo_illustration}
\end{figure*}

\begin{figure*}[!ht]
  \centering
  \subfloat[]{\includegraphics[width=0.45\textwidth]{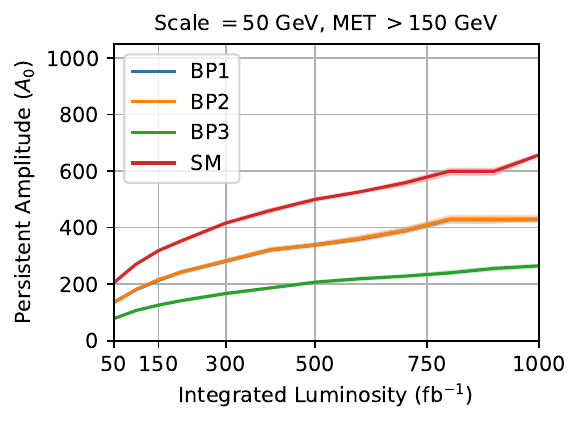}}
  \subfloat[]{\includegraphics[width=0.45\textwidth]{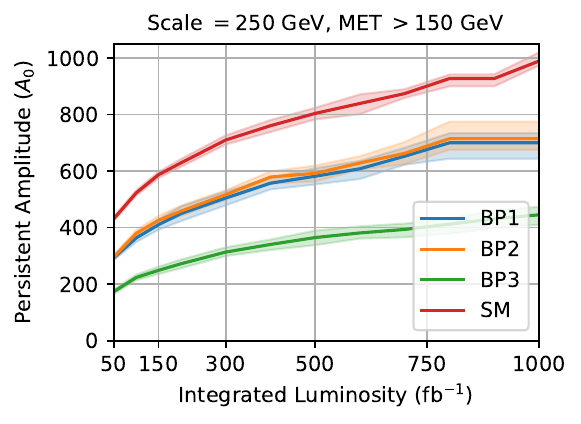}}
  \vskip 0.1in
  \subfloat[]{\includegraphics[width=0.45\textwidth]{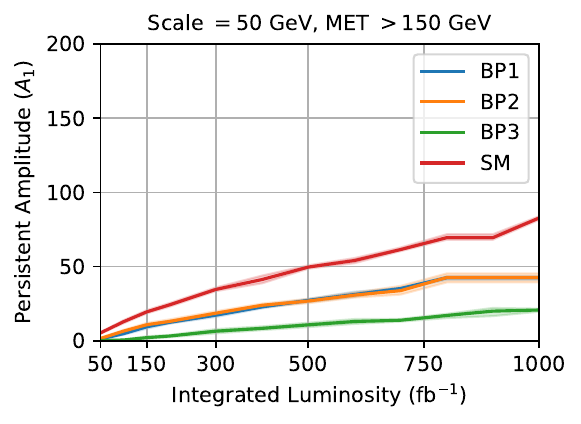}}
  \subfloat[]{\includegraphics[width=0.45\textwidth]{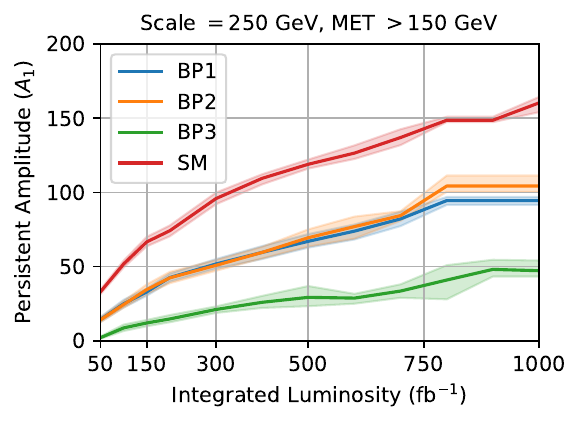}}
  \caption{Persistent Amplitude for two scales (50 GeV and 250 GeV) and MET $> 150$ GeV across different values of integrated luminosity are shown. The shaded regions correspond to the 95\% CL. The first (second) row corresponds to the zeroth (first) homology group.}
  \label{fig:pa_lums}
\end{figure*}

In Fig. \ref{fig:acc_R1_IL150_lums}, we present the classification of Ricci curvature distributions for the SM and the BSM processes across different values of integrated luminosity. In the traditional cut-and-count approach, the significance of discrimination increases with integrated luminosity. However, it is not very obvious for $R_1$ distribution-based classification. We have chosen the benchmark points BP1 and BP2, already excluded by BSM searches at the LHC. This is corroborated by very high classification accuracy, even at a low value of integrated luminosity. Also, the benchmark point BP3 attains appreciable accuracy, particularly for MET $>150$ GeV. However, Table \ref{tab:xsection} shows that BP3 is not excluded when traditional cut-and-count analysis is employed. 

The discussion of Ricci curvature entails the local topological properties of the LHC observations. Next, we delve into the discussion of persistent homology that characterises the global geometrical properties of the LHC events. 

\begin{figure*}[!ht]
  \centering
  \subfloat[]{\includegraphics[width=0.45\textwidth]{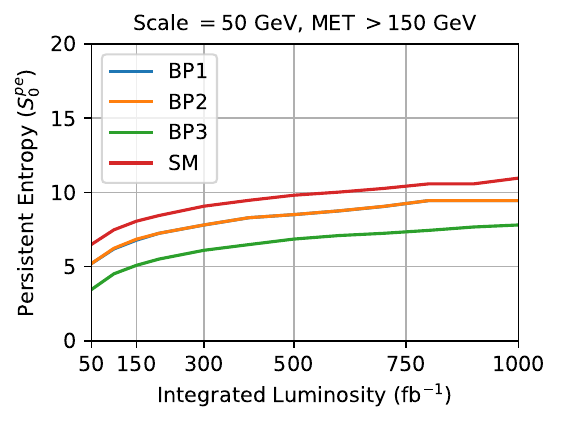}}
  \subfloat[]{\includegraphics[width=0.45\textwidth]{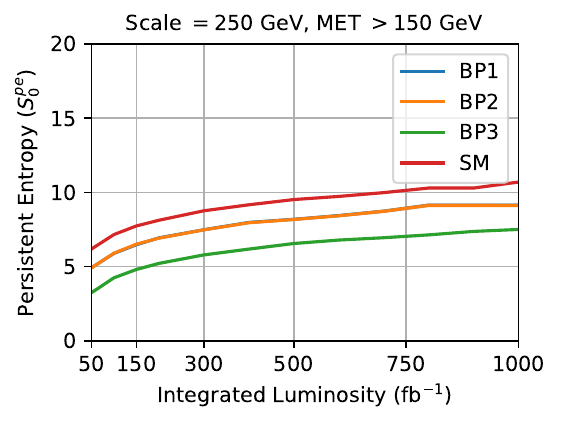}}
  \vskip 0.1in
  \subfloat[]{\includegraphics[width=0.45\textwidth]{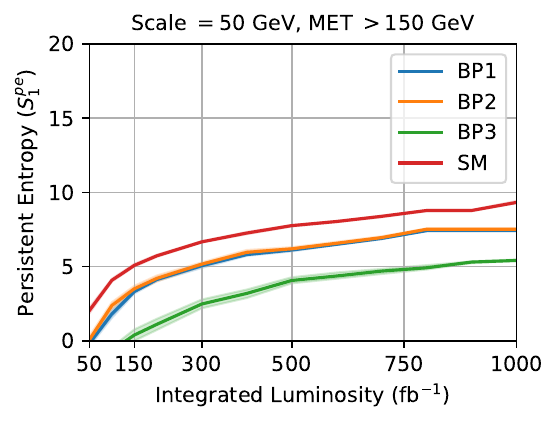}}
  \subfloat[]{\includegraphics[width=0.45\textwidth]{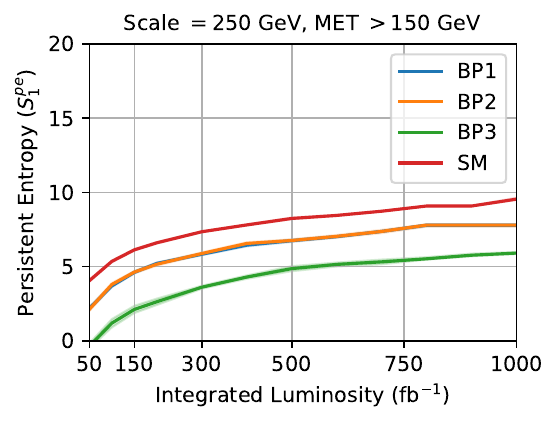}}
  \caption{Persistent entropy for two scales (50 GeV and 250 GeV) and MET $> 150$ GeV across different values of integrated luminosity are shown. The shaded regions correspond to the 95\% CL. The first (second) row corresponds to the zeroth (first) homology group.}
  \label{fig:pe_lums}
\end{figure*}

\begin{figure*}[!ht]
  \centering
  \subfloat[]{\includegraphics[width=0.45\textwidth]{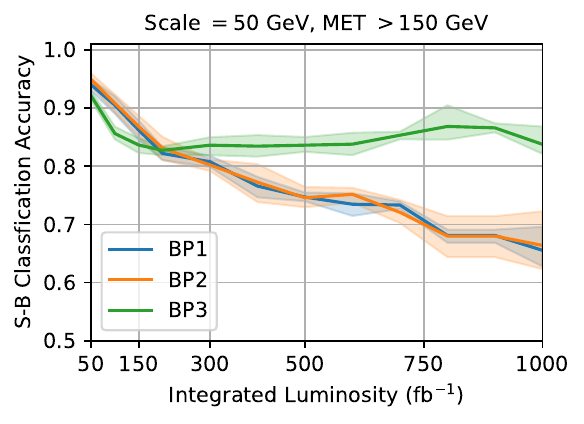}}
  \subfloat[]{\includegraphics[width=0.45\textwidth]{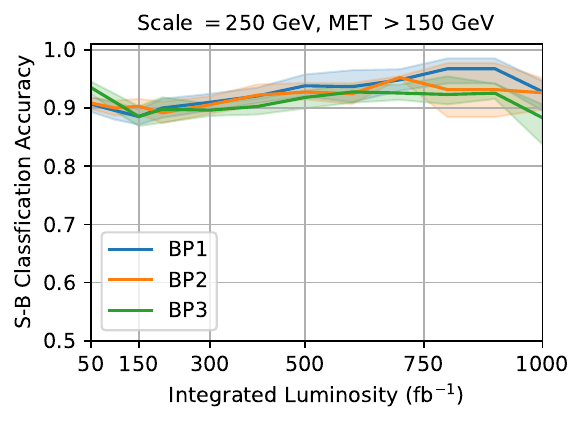}}
  \caption{Signal-Background classification accuracy for two scales (50 GeV and 250 GeV) at MET $> 150$ GeV across different values of integrated luminosity. The shaded regions correspond to the 95\% CL.}
  \label{fig:acc_ph}
\end{figure*}

\subsection{Persistent Homology features}
Before we delve into the discussion of the persistent homology of collider observations, we first illustrate some toy examples. In Fig. \ref{fig:demo_illustration}, we present two distributions $D_1$ and $D_2$ sampled from an annular patch (first row) and an elliptical patch (second row). We vary the number of points from 100 to 2500 for each distribution. We use a maximum filtration scale of 0.2 to form the Vietoris-Rips complex. We find that as the number of sampled points increases, topological features such as persistent entropy and persistent amplitude attain constant values. This is expected because topology is the study of global features of the dataset.

The persistent amplitude is shown in Fig. \ref{fig:demo_illustration}(b) and Fig. \ref{fig:demo_illustration}(d). We observe that it is almost featureless for the elliptic patch. This is because the annular patch has a hole that remains topologically invariant. In other words, the hole cannot be removed without tearing apart the annulus. As shown in Fig. \ref{fig:demo_illustration}(a), $D_2$ has a smaller inner radius and larger outer radius compared to $D_1$. This results in a larger persistent amplitude for $D_2$ than $D_1$ across zeroth and first homology dimensions. The situation remains the same for persistent entropy shown in Fig. \ref{fig:demo_illustration}(c) and Fig. \ref{fig:demo_illustration}(f). Persistent entropy for the zeroth homology group $S_0^{pe}$ remains the same for $D_1$ and  $D_2$ in the case of both annular and elliptic patches. However, as mentioned before, topological properties of $D_1$ and $D_2$ attain a constant ratio with rise in number of points. We will also observe similar behaviour in our latter analysis of LHC events while comparing distributions of the SM background and the BSM scenarios.

Now, we return to the discussion on collider observations. As mentioned before, we use unweighted simplicial complexes to study the persistent homology of the SM background and the benchmark scenarios. The qualitative nature of the conclusions presented here will not change even if we assign weights to edges and vertices in the simplicial complex. In Fig. \ref{fig:pa_lums}, we present the variation of persistent amplitude ($A_0$ and $A_1$) as integrated luminosity increases for two scales of maximum filtration (50 GeV and 250 GeV) while keeping MET$ > $150 GeV. 

As noted in the above-mentioned illustrative example, $A_0$ and $A_1$ for the SM background and the benchmark scenarios attain an approximate constant scaling. The SM background always features a larger magnitude of amplitude because it contributes a larger number of vertices to the simplicial complex. We find that the scale of maximum filtration does have an impact on the persistent amplitude. A scale of 250 GeV favours larger $A_0$ and $A_1$ than 50 GeV scale. This can possibly be attributed to the inclusion of more edge connectivity in the simplicial complex. 

In Fig. \ref{fig:pe_lums}, we present the variation of persistent entropy ($S_0^{pe}$ and $S_1^{pe}$) as integrated luminosity increases. The variation across luminosity is similar to Fig. \ref{fig:pa_lums}. While $S_0^{pe}$ and $S_1^{pe}$ for BP1 and BP2 behave almost similarly,  the ratio with the SM background attains an almost constant scaling across luminosity. The situation is slightly different for BP3, and persistent entropy for BP3 is always the lowest. This is mostly because of its lower cross-section than the other benchmark points.

It is important to note that this scaling of topological properties does not tell us much about the signal-background discrimination potential of persistent homology. We show that the Betti number distribution of the SM process is different from that of the BSM processes. We use Betti number distributions across filtration parameters at a particular scale to train the SVM classifier. The variation of the classification accuracy with integrated luminosity at two different scales of maximum filtration (50 GeV and 250 GeV) while keeping MET $>150$ GeV is presented in Fig. \ref{fig:acc_ph}. 

We notice that the signal-background classification accuracy drops for BP1 and BP2 as integrated luminosity increases when the scale of maximum filtration is 50 GeV. Interestingly, the curve for BP3 remains almost constant. We also notice that the classification accuracy remains almost constant for all benchmark points as integrated luminosity increases for the 250 GeV scale. Thus, we can conclude that MET $>150$ GeV likely prefers a higher scale of maximum filtration. This observation is crucial and conveys the importance of the suitable scale while studying the topological properties of collider observations.

\section{Conclusion}
\label{sec:conclusion}

In conclusion, our study has provided insights into the distinction between the topological properties of the BSM signal and the SM background events. The framework discussed in this work is quite generic and can be applied to other BSM searches at colliders. Through the analysis of Forman Ricci curvature and persistent homology, we have explored the dependence of topological features on kinematic cut selection, integrated luminosity, and the scale of maximum filtration.

Our results highlight the efficacy of topological properties, such as Forman Ricci curvature and persistent homology, in capturing intrinsic geometric structures present in collider observations. This study also emphasizes the impact of MET cuts and the scale of maximum filtration while studying topological properties. Furthermore, our investigation into persistent homology features reveals an approximate scaling behaviour across integrated luminosity, suggesting robustness in topological properties irrespective of the dataset size. 

Crucially, our study underscores the interconnectedness between the scale of maximum filtration, MET cuts, and the mass of the LSP, highlighting the importance of considering these factors collectively in topological analyses aimed at uncovering new physics scenarios. We believe using topological discriminators with a traditional kinematic cut-based approach can yield stronger exclusion limits even at low integrated luminosity. 

Overall, our findings deepen the understanding of the intrinsic geometric structures in collider observations and provide valuable implications for future experimental studies to uncover novel physics phenomena beyond the Standard Model.

\section*{Supplementary Material}
The Python code used in this work is available at \url{https://github.com/jbeuria/FRC-PH-LHC/}.

\section*{ACKNOWLEDGMENTS}
JB thanks IKSMHA Center, IIT Mandi and IKS Research Center, ISS Delhi, for their kind support.

\bibliography{ref}

\providecommand{\noopsort}[1]{}\providecommand{\singleletter}[1]{#1}%
\begin{thebibliography}{58}%
\makeatletter
\providecommand \@ifxundefined [1]{%
 \@ifx{#1\undefined}
}%
\providecommand \@ifnum [1]{%
 \ifnum #1\expandafter \@firstoftwo
 \else \expandafter \@secondoftwo
 \fi
}%
\providecommand \@ifx [1]{%
 \ifx #1\expandafter \@firstoftwo
 \else \expandafter \@secondoftwo
 \fi
}%
\providecommand \natexlab [1]{#1}%
\providecommand \enquote  [1]{``#1''}%
\providecommand \bibnamefont  [1]{#1}%
\providecommand \bibfnamefont [1]{#1}%
\providecommand \citenamefont [1]{#1}%
\providecommand \href@noop [0]{\@secondoftwo}%
\providecommand \href [0]{\begingroup \@sanitize@url \@href}%
\providecommand \@href[1]{\@@startlink{#1}\@@href}%
\providecommand \@@href[1]{\endgroup#1\@@endlink}%
\providecommand \@sanitize@url [0]{\catcode `\\12\catcode `\$12\catcode
  `\&12\catcode `\#12\catcode `\^12\catcode `\_12\catcode `\%12\relax}%
\providecommand \@@startlink[1]{}%
\providecommand \@@endlink[0]{}%
\providecommand \url  [0]{\begingroup\@sanitize@url \@url }%
\providecommand \@url [1]{\endgroup\@href {#1}{\urlprefix }}%
\providecommand \urlprefix  [0]{URL }%
\providecommand \Eprint [0]{\href }%
\providecommand \doibase [0]{https://doi.org/}%
\providecommand \selectlanguage [0]{\@gobble}%
\providecommand \bibinfo  [0]{\@secondoftwo}%
\providecommand \bibfield  [0]{\@secondoftwo}%
\providecommand \translation [1]{[#1]}%
\providecommand \BibitemOpen [0]{}%
\providecommand \bibitemStop [0]{}%
\providecommand \bibitemNoStop [0]{.\EOS\space}%
\providecommand \EOS [0]{\spacefactor3000\relax}%
\providecommand \BibitemShut  [1]{\csname bibitem#1\endcsname}%
\let\auto@bib@innerbib\@empty
\bibitem [{\citenamefont {Aad}\ \emph {et~al.}(2012)\citenamefont {Aad},
  \citenamefont {Abajyan}, \citenamefont {Abbott}, \citenamefont {Abdallah},
  \citenamefont {Khalek}, \citenamefont {Abdelalim}, \citenamefont {Aben},
  \citenamefont {Abi}, \citenamefont {Abolins}, \citenamefont {AbouZeid} \emph
  {et~al.}}]{aad2012observation}%
  \BibitemOpen
  \bibfield  {author} {\bibinfo {author} {\bibfnamefont {G.}~\bibnamefont
  {Aad}}, \bibinfo {author} {\bibfnamefont {T.}~\bibnamefont {Abajyan}},
  \bibinfo {author} {\bibfnamefont {B.}~\bibnamefont {Abbott}}, \bibinfo
  {author} {\bibfnamefont {J.}~\bibnamefont {Abdallah}}, \bibinfo {author}
  {\bibfnamefont {S.~A.}\ \bibnamefont {Khalek}}, \bibinfo {author}
  {\bibfnamefont {A.~A.}\ \bibnamefont {Abdelalim}}, \bibinfo {author}
  {\bibfnamefont {R.}~\bibnamefont {Aben}}, \bibinfo {author} {\bibfnamefont
  {B.}~\bibnamefont {Abi}}, \bibinfo {author} {\bibfnamefont {M.}~\bibnamefont
  {Abolins}}, \bibinfo {author} {\bibfnamefont {O.}~\bibnamefont {AbouZeid}},
  \emph {et~al.},\ }\bibfield  {title} {\bibinfo {title} {Observation of a new
  particle in the search for the standard model higgs boson with the atlas
  detector at the lhc},\ }\href@noop {} {\bibfield  {journal} {\bibinfo
  {journal} {Physics Letters B}\ }\textbf {\bibinfo {volume} {716}},\ \bibinfo
  {pages} {1} (\bibinfo {year} {2012})}\BibitemShut {NoStop}%
\bibitem [{\citenamefont {Chatrchyan}\ \emph {et~al.}(2012)\citenamefont
  {Chatrchyan}, \citenamefont {Khachatryan}, \citenamefont {Sirunyan},
  \citenamefont {Tumasyan}, \citenamefont {Adam}, \citenamefont {Aguilo},
  \citenamefont {Bergauer}, \citenamefont {Dragicevic}, \citenamefont
  {Er{\"o}}, \citenamefont {Fabjan} \emph
  {et~al.}}]{chatrchyan2012observation}%
  \BibitemOpen
  \bibfield  {author} {\bibinfo {author} {\bibfnamefont {S.}~\bibnamefont
  {Chatrchyan}}, \bibinfo {author} {\bibfnamefont {V.}~\bibnamefont
  {Khachatryan}}, \bibinfo {author} {\bibfnamefont {A.~M.}\ \bibnamefont
  {Sirunyan}}, \bibinfo {author} {\bibfnamefont {A.}~\bibnamefont {Tumasyan}},
  \bibinfo {author} {\bibfnamefont {W.}~\bibnamefont {Adam}}, \bibinfo {author}
  {\bibfnamefont {E.}~\bibnamefont {Aguilo}}, \bibinfo {author} {\bibfnamefont
  {T.}~\bibnamefont {Bergauer}}, \bibinfo {author} {\bibfnamefont
  {M.}~\bibnamefont {Dragicevic}}, \bibinfo {author} {\bibfnamefont
  {J.}~\bibnamefont {Er{\"o}}}, \bibinfo {author} {\bibfnamefont
  {C.}~\bibnamefont {Fabjan}}, \emph {et~al.},\ }\bibfield  {title} {\bibinfo
  {title} {Observation of a new boson at a mass of 125 gev with the cms
  experiment at the lhc},\ }\href@noop {} {\bibfield  {journal} {\bibinfo
  {journal} {Physics Letters B}\ }\textbf {\bibinfo {volume} {716}},\ \bibinfo
  {pages} {30} (\bibinfo {year} {2012})}\BibitemShut {NoStop}%
\bibitem [{\citenamefont {Franceschini}\ \emph {et~al.}(2023)\citenamefont
  {Franceschini}, \citenamefont {Kim}, \citenamefont {Kong}, \citenamefont
  {Matchev}, \citenamefont {Park},\ and\ \citenamefont
  {Shyamsundar}}]{franceschini2023kinematic}%
  \BibitemOpen
  \bibfield  {author} {\bibinfo {author} {\bibfnamefont {R.}~\bibnamefont
  {Franceschini}}, \bibinfo {author} {\bibfnamefont {D.}~\bibnamefont {Kim}},
  \bibinfo {author} {\bibfnamefont {K.}~\bibnamefont {Kong}}, \bibinfo {author}
  {\bibfnamefont {K.~T.}\ \bibnamefont {Matchev}}, \bibinfo {author}
  {\bibfnamefont {M.}~\bibnamefont {Park}},\ and\ \bibinfo {author}
  {\bibfnamefont {P.}~\bibnamefont {Shyamsundar}},\ }\href@noop {} {\bibinfo
  {title} {Kinematic variables and feature engineering for particle
  phenomenology}} (\bibinfo {year} {2023})\BibitemShut {NoStop}%
\bibitem [{\citenamefont {Debnath}\ \emph
  {et~al.}(2016{\natexlab{a}})\citenamefont {Debnath}, \citenamefont {Gainer},
  \citenamefont {Kim},\ and\ \citenamefont {Matchev}}]{Debnath:2015wra}%
  \BibitemOpen
  \bibfield  {author} {\bibinfo {author} {\bibfnamefont {D.}~\bibnamefont
  {Debnath}}, \bibinfo {author} {\bibfnamefont {J.~S.}\ \bibnamefont {Gainer}},
  \bibinfo {author} {\bibfnamefont {D.}~\bibnamefont {Kim}},\ and\ \bibinfo
  {author} {\bibfnamefont {K.~T.}\ \bibnamefont {Matchev}},\ }\bibfield
  {title} {\bibinfo {title} {{Edge Detecting New Physics the Voronoi Way}},\
  }\href {https://doi.org/10.1209/0295-5075/114/41001} {\bibfield  {journal}
  {\bibinfo  {journal} {EPL}\ }\textbf {\bibinfo {volume} {114}},\ \bibinfo
  {pages} {41001} (\bibinfo {year} {2016}{\natexlab{a}})},\ \Eprint
  {https://arxiv.org/abs/1506.04141} {arXiv:1506.04141 [hep-ph]} \BibitemShut
  {NoStop}%
\bibitem [{\citenamefont {Debnath}\ \emph
  {et~al.}(2016{\natexlab{b}})\citenamefont {Debnath}, \citenamefont {Gainer},
  \citenamefont {Kilic}, \citenamefont {Kim}, \citenamefont {Matchev},\ and\
  \citenamefont {Yang}}]{Debnath:2016mwb}%
  \BibitemOpen
  \bibfield  {author} {\bibinfo {author} {\bibfnamefont {D.}~\bibnamefont
  {Debnath}}, \bibinfo {author} {\bibfnamefont {J.~S.}\ \bibnamefont {Gainer}},
  \bibinfo {author} {\bibfnamefont {C.}~\bibnamefont {Kilic}}, \bibinfo
  {author} {\bibfnamefont {D.}~\bibnamefont {Kim}}, \bibinfo {author}
  {\bibfnamefont {K.~T.}\ \bibnamefont {Matchev}},\ and\ \bibinfo {author}
  {\bibfnamefont {Y.-P.}\ \bibnamefont {Yang}},\ }\bibfield  {title} {\bibinfo
  {title} {{Identifying Phase Space Boundaries with Voronoi Tessellations}},\
  }\href {https://doi.org/10.1140/epjc/s10052-016-4431-z} {\bibfield  {journal}
  {\bibinfo  {journal} {Eur. Phys. J. C}\ }\textbf {\bibinfo {volume} {76}},\
  \bibinfo {pages} {645} (\bibinfo {year} {2016}{\natexlab{b}})},\ \Eprint
  {https://arxiv.org/abs/1606.02721} {arXiv:1606.02721 [hep-ph]} \BibitemShut
  {NoStop}%
\bibitem [{\citenamefont {Matchev}\ \emph {et~al.}(2020)\citenamefont
  {Matchev}, \citenamefont {Roman},\ and\ \citenamefont
  {Shyamsundar}}]{Matchev:2020vhr}%
  \BibitemOpen
  \bibfield  {author} {\bibinfo {author} {\bibfnamefont {K.~T.}\ \bibnamefont
  {Matchev}}, \bibinfo {author} {\bibfnamefont {A.}~\bibnamefont {Roman}},\
  and\ \bibinfo {author} {\bibfnamefont {P.}~\bibnamefont {Shyamsundar}},\
  }\bibfield  {title} {\bibinfo {title} {{Finding wombling boundaries in LHC
  data with Voronoi and Delaunay tessellations}},\ }\href
  {https://doi.org/10.1007/JHEP12(2020)137} {\bibfield  {journal} {\bibinfo
  {journal} {JHEP}\ }\textbf {\bibinfo {volume} {12}},\ \bibinfo {pages}
  {137}},\ \Eprint {https://arxiv.org/abs/2006.06582} {arXiv:2006.06582
  [hep-ph]} \BibitemShut {NoStop}%
\bibitem [{\citenamefont {Mullin}\ \emph {et~al.}(2021)\citenamefont {Mullin},
  \citenamefont {Nicholls}, \citenamefont {Pacey}, \citenamefont {Parker},
  \citenamefont {White},\ and\ \citenamefont {Williams}}]{Mullin:2019mmh}%
  \BibitemOpen
  \bibfield  {author} {\bibinfo {author} {\bibfnamefont {A.}~\bibnamefont
  {Mullin}}, \bibinfo {author} {\bibfnamefont {S.}~\bibnamefont {Nicholls}},
  \bibinfo {author} {\bibfnamefont {H.}~\bibnamefont {Pacey}}, \bibinfo
  {author} {\bibfnamefont {M.}~\bibnamefont {Parker}}, \bibinfo {author}
  {\bibfnamefont {M.}~\bibnamefont {White}},\ and\ \bibinfo {author}
  {\bibfnamefont {S.}~\bibnamefont {Williams}},\ }\bibfield  {title} {\bibinfo
  {title} {{Does SUSY have friends? A new approach for LHC event analysis}},\
  }\href {https://doi.org/10.1007/JHEP02(2021)160} {\bibfield  {journal}
  {\bibinfo  {journal} {JHEP}\ }\textbf {\bibinfo {volume} {02}},\ \bibinfo
  {pages} {160}},\ \Eprint {https://arxiv.org/abs/1912.10625} {arXiv:1912.10625
  [hep-ph]} \BibitemShut {NoStop}%
\bibitem [{\citenamefont {Guest}\ \emph {et~al.}(2018)\citenamefont {Guest},
  \citenamefont {Cranmer},\ and\ \citenamefont {Whiteson}}]{guest2018deep}%
  \BibitemOpen
  \bibfield  {author} {\bibinfo {author} {\bibfnamefont {D.}~\bibnamefont
  {Guest}}, \bibinfo {author} {\bibfnamefont {K.}~\bibnamefont {Cranmer}},\
  and\ \bibinfo {author} {\bibfnamefont {D.}~\bibnamefont {Whiteson}},\
  }\bibfield  {title} {\bibinfo {title} {Deep learning and its application to
  lhc physics},\ }\href@noop {} {\bibfield  {journal} {\bibinfo  {journal}
  {Annual Review of Nuclear and Particle Science}\ }\textbf {\bibinfo {volume}
  {68}},\ \bibinfo {pages} {161} (\bibinfo {year} {2018})}\BibitemShut
  {NoStop}%
\bibitem [{\citenamefont {Qasim}\ \emph {et~al.}(2019)\citenamefont {Qasim},
  \citenamefont {Kieseler}, \citenamefont {Iiyama},\ and\ \citenamefont
  {Pierini}}]{qasim2019learning}%
  \BibitemOpen
  \bibfield  {author} {\bibinfo {author} {\bibfnamefont {S.~R.}\ \bibnamefont
  {Qasim}}, \bibinfo {author} {\bibfnamefont {J.}~\bibnamefont {Kieseler}},
  \bibinfo {author} {\bibfnamefont {Y.}~\bibnamefont {Iiyama}},\ and\ \bibinfo
  {author} {\bibfnamefont {M.}~\bibnamefont {Pierini}},\ }\bibfield  {title}
  {\bibinfo {title} {Learning representations of irregular particle-detector
  geometry with distance-weighted graph networks},\ }\href@noop {} {\bibfield
  {journal} {\bibinfo  {journal} {The European Physical Journal C}\ }\textbf
  {\bibinfo {volume} {79}},\ \bibinfo {pages} {1} (\bibinfo {year}
  {2019})}\BibitemShut {NoStop}%
\bibitem [{\citenamefont {Abdughani}\ \emph {et~al.}(2019)\citenamefont
  {Abdughani}, \citenamefont {Ren}, \citenamefont {Wu},\ and\ \citenamefont
  {Yang}}]{abdughani2019probing}%
  \BibitemOpen
  \bibfield  {author} {\bibinfo {author} {\bibfnamefont {M.}~\bibnamefont
  {Abdughani}}, \bibinfo {author} {\bibfnamefont {J.}~\bibnamefont {Ren}},
  \bibinfo {author} {\bibfnamefont {L.}~\bibnamefont {Wu}},\ and\ \bibinfo
  {author} {\bibfnamefont {J.~M.}\ \bibnamefont {Yang}},\ }\bibfield  {title}
  {\bibinfo {title} {Probing stop pair production at the lhc with graph neural
  networks},\ }\href@noop {} {\bibfield  {journal} {\bibinfo  {journal}
  {Journal of High Energy Physics}\ }\textbf {\bibinfo {volume} {2019}},\
  \bibinfo {pages} {1} (\bibinfo {year} {2019})}\BibitemShut {NoStop}%
\bibitem [{\citenamefont {Du}\ \emph {et~al.}(2020)\citenamefont {Du},
  \citenamefont {Zhou}, \citenamefont {Steinheimer}, \citenamefont {Pang},
  \citenamefont {Motornenko}, \citenamefont {Zong}, \citenamefont {Wang},\ and\
  \citenamefont {St{\"o}cker}}]{du2020identifying}%
  \BibitemOpen
  \bibfield  {author} {\bibinfo {author} {\bibfnamefont {Y.-L.}\ \bibnamefont
  {Du}}, \bibinfo {author} {\bibfnamefont {K.}~\bibnamefont {Zhou}}, \bibinfo
  {author} {\bibfnamefont {J.}~\bibnamefont {Steinheimer}}, \bibinfo {author}
  {\bibfnamefont {L.-G.}\ \bibnamefont {Pang}}, \bibinfo {author}
  {\bibfnamefont {A.}~\bibnamefont {Motornenko}}, \bibinfo {author}
  {\bibfnamefont {H.-S.}\ \bibnamefont {Zong}}, \bibinfo {author}
  {\bibfnamefont {X.-N.}\ \bibnamefont {Wang}},\ and\ \bibinfo {author}
  {\bibfnamefont {H.}~\bibnamefont {St{\"o}cker}},\ }\bibfield  {title}
  {\bibinfo {title} {Identifying the nature of the qcd transition in
  relativistic collision of heavy nuclei with deep learning},\ }\href@noop {}
  {\bibfield  {journal} {\bibinfo  {journal} {The European Physical Journal C}\
  }\textbf {\bibinfo {volume} {80}},\ \bibinfo {pages} {1} (\bibinfo {year}
  {2020})}\BibitemShut {NoStop}%
\bibitem [{\citenamefont {Flesher}\ \emph {et~al.}(2021)\citenamefont
  {Flesher}, \citenamefont {Fraser}, \citenamefont {Hutchison}, \citenamefont
  {Ostdiek},\ and\ \citenamefont {Schwartz}}]{flesher2021parameter}%
  \BibitemOpen
  \bibfield  {author} {\bibinfo {author} {\bibfnamefont {F.}~\bibnamefont
  {Flesher}}, \bibinfo {author} {\bibfnamefont {K.}~\bibnamefont {Fraser}},
  \bibinfo {author} {\bibfnamefont {C.}~\bibnamefont {Hutchison}}, \bibinfo
  {author} {\bibfnamefont {B.}~\bibnamefont {Ostdiek}},\ and\ \bibinfo {author}
  {\bibfnamefont {M.~D.}\ \bibnamefont {Schwartz}},\ }\bibfield  {title}
  {\bibinfo {title} {Parameter inference from event ensembles and the top-quark
  mass},\ }\href@noop {} {\bibfield  {journal} {\bibinfo  {journal} {Journal of
  High Energy Physics}\ }\textbf {\bibinfo {volume} {2021}},\ \bibinfo {pages}
  {1} (\bibinfo {year} {2021})}\BibitemShut {NoStop}%
\bibitem [{\citenamefont {Chang}\ \emph {et~al.}(2021)\citenamefont {Chang},
  \citenamefont {Chen},\ and\ \citenamefont {Chiang}}]{Chang:2020rtc}%
  \BibitemOpen
  \bibfield  {author} {\bibinfo {author} {\bibfnamefont {S.}~\bibnamefont
  {Chang}}, \bibinfo {author} {\bibfnamefont {T.-K.}\ \bibnamefont {Chen}},\
  and\ \bibinfo {author} {\bibfnamefont {C.-W.}\ \bibnamefont {Chiang}},\
  }\bibfield  {title} {\bibinfo {title} {Distinguishing ${W}^{\ensuremath{'}}$
  signals at hadron colliders using neural networks},\ }\href
  {https://doi.org/10.1103/PhysRevD.103.036016} {\bibfield  {journal} {\bibinfo
   {journal} {Phys. Rev. D}\ }\textbf {\bibinfo {volume} {103}},\ \bibinfo
  {pages} {036016} (\bibinfo {year} {2021})}\BibitemShut {NoStop}%
\bibitem [{\citenamefont {Nachman}\ and\ \citenamefont
  {Thaler}(2021)}]{Nachman:2021yvi}%
  \BibitemOpen
  \bibfield  {author} {\bibinfo {author} {\bibfnamefont {B.}~\bibnamefont
  {Nachman}}\ and\ \bibinfo {author} {\bibfnamefont {J.}~\bibnamefont
  {Thaler}},\ }\bibfield  {title} {\bibinfo {title} {{Learning from many
  collider events at once}},\ }\href
  {https://doi.org/10.1103/PhysRevD.103.116013} {\bibfield  {journal} {\bibinfo
   {journal} {Phys. Rev. D}\ }\textbf {\bibinfo {volume} {103}},\ \bibinfo
  {pages} {116013} (\bibinfo {year} {2021})},\ \Eprint
  {https://arxiv.org/abs/2101.07263} {arXiv:2101.07263 [physics.data-an]}
  \BibitemShut {NoStop}%
\bibitem [{\citenamefont {Knipfer}\ \emph {et~al.}(2023)\citenamefont
  {Knipfer}, \citenamefont {Meier}, \citenamefont {Heimerl}, \citenamefont
  {Hommelhoff},\ and\ \citenamefont {Gleyzer}}]{knipfer2023deep}%
  \BibitemOpen
  \bibfield  {author} {\bibinfo {author} {\bibfnamefont {M.}~\bibnamefont
  {Knipfer}}, \bibinfo {author} {\bibfnamefont {S.}~\bibnamefont {Meier}},
  \bibinfo {author} {\bibfnamefont {J.}~\bibnamefont {Heimerl}}, \bibinfo
  {author} {\bibfnamefont {P.}~\bibnamefont {Hommelhoff}},\ and\ \bibinfo
  {author} {\bibfnamefont {S.}~\bibnamefont {Gleyzer}},\ }\bibfield  {title}
  {\bibinfo {title} {Deep learning-based spatiotemporal multi-event
  reconstruction for delay line detectors},\ }\href@noop {} {\bibfield
  {journal} {\bibinfo  {journal} {arXiv preprint arXiv:2306.09359}\ } (\bibinfo
  {year} {2023})}\BibitemShut {NoStop}%
\bibitem [{\citenamefont {Taylor}\ \emph {et~al.}(2015)\citenamefont {Taylor},
  \citenamefont {Klimm}, \citenamefont {Harrington}, \citenamefont
  {Kram{\'a}r}, \citenamefont {Mischaikow}, \citenamefont {Porter},\ and\
  \citenamefont {Mucha}}]{taylor2015topological}%
  \BibitemOpen
  \bibfield  {author} {\bibinfo {author} {\bibfnamefont {D.}~\bibnamefont
  {Taylor}}, \bibinfo {author} {\bibfnamefont {F.}~\bibnamefont {Klimm}},
  \bibinfo {author} {\bibfnamefont {H.~A.}\ \bibnamefont {Harrington}},
  \bibinfo {author} {\bibfnamefont {M.}~\bibnamefont {Kram{\'a}r}}, \bibinfo
  {author} {\bibfnamefont {K.}~\bibnamefont {Mischaikow}}, \bibinfo {author}
  {\bibfnamefont {M.~A.}\ \bibnamefont {Porter}},\ and\ \bibinfo {author}
  {\bibfnamefont {P.~J.}\ \bibnamefont {Mucha}},\ }\bibfield  {title} {\bibinfo
  {title} {Topological data analysis of contagion maps for examining spreading
  processes on networks},\ }\href@noop {} {\bibfield  {journal} {\bibinfo
  {journal} {Nature communications}\ }\textbf {\bibinfo {volume} {6}},\
  \bibinfo {pages} {7723} (\bibinfo {year} {2015})}\BibitemShut {NoStop}%
\bibitem [{\citenamefont {Topaz}\ \emph {et~al.}(2015)\citenamefont {Topaz},
  \citenamefont {Ziegelmeier},\ and\ \citenamefont
  {Halverson}}]{topaz2015topological}%
  \BibitemOpen
  \bibfield  {author} {\bibinfo {author} {\bibfnamefont {C.~M.}\ \bibnamefont
  {Topaz}}, \bibinfo {author} {\bibfnamefont {L.}~\bibnamefont {Ziegelmeier}},\
  and\ \bibinfo {author} {\bibfnamefont {T.}~\bibnamefont {Halverson}},\
  }\bibfield  {title} {\bibinfo {title} {Topological data analysis of
  biological aggregation models},\ }\href@noop {} {\bibfield  {journal}
  {\bibinfo  {journal} {PloS one}\ }\textbf {\bibinfo {volume} {10}},\ \bibinfo
  {pages} {e0126383} (\bibinfo {year} {2015})}\BibitemShut {NoStop}%
\bibitem [{\citenamefont {Lloyd}\ \emph {et~al.}(2016)\citenamefont {Lloyd},
  \citenamefont {Garnerone},\ and\ \citenamefont {Zanardi}}]{lloyd2016quantum}%
  \BibitemOpen
  \bibfield  {author} {\bibinfo {author} {\bibfnamefont {S.}~\bibnamefont
  {Lloyd}}, \bibinfo {author} {\bibfnamefont {S.}~\bibnamefont {Garnerone}},\
  and\ \bibinfo {author} {\bibfnamefont {P.}~\bibnamefont {Zanardi}},\
  }\bibfield  {title} {\bibinfo {title} {Quantum algorithms for topological and
  geometric analysis of data},\ }\href@noop {} {\bibfield  {journal} {\bibinfo
  {journal} {Nature communications}\ }\textbf {\bibinfo {volume} {7}},\
  \bibinfo {pages} {10138} (\bibinfo {year} {2016})}\BibitemShut {NoStop}%
\bibitem [{\citenamefont {Gidea}\ and\ \citenamefont
  {Katz}(2018)}]{gidea2018topological}%
  \BibitemOpen
  \bibfield  {author} {\bibinfo {author} {\bibfnamefont {M.}~\bibnamefont
  {Gidea}}\ and\ \bibinfo {author} {\bibfnamefont {Y.}~\bibnamefont {Katz}},\
  }\bibfield  {title} {\bibinfo {title} {Topological data analysis of financial
  time series: Landscapes of crashes},\ }\href@noop {} {\bibfield  {journal}
  {\bibinfo  {journal} {Physica A: Statistical Mechanics and its Applications}\
  }\textbf {\bibinfo {volume} {491}},\ \bibinfo {pages} {820} (\bibinfo {year}
  {2018})}\BibitemShut {NoStop}%
\bibitem [{\citenamefont {Saggar}\ \emph {et~al.}(2018)\citenamefont {Saggar},
  \citenamefont {Sporns}, \citenamefont {Gonzalez-Castillo}, \citenamefont
  {Bandettini}, \citenamefont {Carlsson}, \citenamefont {Glover},\ and\
  \citenamefont {Reiss}}]{saggar2018towards}%
  \BibitemOpen
  \bibfield  {author} {\bibinfo {author} {\bibfnamefont {M.}~\bibnamefont
  {Saggar}}, \bibinfo {author} {\bibfnamefont {O.}~\bibnamefont {Sporns}},
  \bibinfo {author} {\bibfnamefont {J.}~\bibnamefont {Gonzalez-Castillo}},
  \bibinfo {author} {\bibfnamefont {P.~A.}\ \bibnamefont {Bandettini}},
  \bibinfo {author} {\bibfnamefont {G.}~\bibnamefont {Carlsson}}, \bibinfo
  {author} {\bibfnamefont {G.}~\bibnamefont {Glover}},\ and\ \bibinfo {author}
  {\bibfnamefont {A.~L.}\ \bibnamefont {Reiss}},\ }\bibfield  {title} {\bibinfo
  {title} {Towards a new approach to reveal dynamical organization of the brain
  using topological data analysis},\ }\href@noop {} {\bibfield  {journal}
  {\bibinfo  {journal} {Nature communications}\ }\textbf {\bibinfo {volume}
  {9}},\ \bibinfo {pages} {1399} (\bibinfo {year} {2018})}\BibitemShut
  {NoStop}%
\bibitem [{\citenamefont {Sizemore}\ \emph {et~al.}(2019)\citenamefont
  {Sizemore}, \citenamefont {Phillips-Cremins}, \citenamefont {Ghrist},\ and\
  \citenamefont {Bassett}}]{sizemore2019importance}%
  \BibitemOpen
  \bibfield  {author} {\bibinfo {author} {\bibfnamefont {A.~E.}\ \bibnamefont
  {Sizemore}}, \bibinfo {author} {\bibfnamefont {J.~E.}\ \bibnamefont
  {Phillips-Cremins}}, \bibinfo {author} {\bibfnamefont {R.}~\bibnamefont
  {Ghrist}},\ and\ \bibinfo {author} {\bibfnamefont {D.~S.}\ \bibnamefont
  {Bassett}},\ }\bibfield  {title} {\bibinfo {title} {The importance of the
  whole: topological data analysis for the network neuroscientist},\
  }\href@noop {} {\bibfield  {journal} {\bibinfo  {journal} {Network
  Neuroscience}\ }\textbf {\bibinfo {volume} {3}},\ \bibinfo {pages} {656}
  (\bibinfo {year} {2019})}\BibitemShut {NoStop}%
\bibitem [{\citenamefont {Murugan}\ and\ \citenamefont
  {Robertson}(2019)}]{murugan2019introduction}%
  \BibitemOpen
  \bibfield  {author} {\bibinfo {author} {\bibfnamefont {J.}~\bibnamefont
  {Murugan}}\ and\ \bibinfo {author} {\bibfnamefont {D.}~\bibnamefont
  {Robertson}},\ }\bibfield  {title} {\bibinfo {title} {An introduction to
  topological data analysis for physicists: From lgm to frbs},\ }\href@noop {}
  {\bibfield  {journal} {\bibinfo  {journal} {arXiv preprint arXiv:1904.11044}\
  } (\bibinfo {year} {2019})}\BibitemShut {NoStop}%
\bibitem [{\citenamefont {Cole}\ and\ \citenamefont
  {Shiu}(2019)}]{cole2019topological}%
  \BibitemOpen
  \bibfield  {author} {\bibinfo {author} {\bibfnamefont {A.}~\bibnamefont
  {Cole}}\ and\ \bibinfo {author} {\bibfnamefont {G.}~\bibnamefont {Shiu}},\
  }\bibfield  {title} {\bibinfo {title} {Topological data analysis for the
  string landscape},\ }\href@noop {} {\bibfield  {journal} {\bibinfo  {journal}
  {Journal of High Energy Physics}\ }\textbf {\bibinfo {volume} {2019}},\
  \bibinfo {pages} {1} (\bibinfo {year} {2019})}\BibitemShut {NoStop}%
\bibitem [{\citenamefont {Chazal}\ and\ \citenamefont
  {Michel}(2021)}]{chazal2021introduction}%
  \BibitemOpen
  \bibfield  {author} {\bibinfo {author} {\bibfnamefont {F.}~\bibnamefont
  {Chazal}}\ and\ \bibinfo {author} {\bibfnamefont {B.}~\bibnamefont
  {Michel}},\ }\bibfield  {title} {\bibinfo {title} {An introduction to
  topological data analysis: fundamental and practical aspects for data
  scientists},\ }\href@noop {} {\bibfield  {journal} {\bibinfo  {journal}
  {Frontiers in artificial intelligence}\ }\textbf {\bibinfo {volume} {4}},\
  \bibinfo {pages} {108} (\bibinfo {year} {2021})}\BibitemShut {NoStop}%
\bibitem [{\citenamefont {Beuria}(2023)}]{beuria2023persistent}%
  \BibitemOpen
  \bibfield  {author} {\bibinfo {author} {\bibfnamefont {J.}~\bibnamefont
  {Beuria}},\ }\bibfield  {title} {\bibinfo {title} {Persistent homology of
  collider observations: When (w) hole matters},\ }\href@noop {} {\bibfield
  {journal} {\bibinfo  {journal} {Physics Letters B}\ }\textbf {\bibinfo
  {volume} {846}},\ \bibinfo {pages} {138188} (\bibinfo {year}
  {2023})}\BibitemShut {NoStop}%
\bibitem [{\citenamefont {Gupta}\ \emph {et~al.}(2024)\citenamefont {Gupta},
  \citenamefont {Beuria},\ and\ \citenamefont
  {Behera}}]{gupta2024characterizing}%
  \BibitemOpen
  \bibfield  {author} {\bibinfo {author} {\bibfnamefont {K.~V.}\ \bibnamefont
  {Gupta}}, \bibinfo {author} {\bibfnamefont {J.}~\bibnamefont {Beuria}},\ and\
  \bibinfo {author} {\bibfnamefont {L.}~\bibnamefont {Behera}},\ }\bibfield
  {title} {\bibinfo {title} {Characterizing eeg signals of meditative states
  using persistent homology and hodge spectral entropy},\ }\href@noop {}
  {\bibfield  {journal} {\bibinfo  {journal} {Biomedical Signal Processing and
  Control}\ }\textbf {\bibinfo {volume} {89}},\ \bibinfo {pages} {105779}
  (\bibinfo {year} {2024})}\BibitemShut {NoStop}%
\bibitem [{\citenamefont {Perelman}(2003)}]{perelman2003ricci}%
  \BibitemOpen
  \bibfield  {author} {\bibinfo {author} {\bibfnamefont {G.}~\bibnamefont
  {Perelman}},\ }\bibfield  {title} {\bibinfo {title} {Ricci flow with surgery
  on three-manifolds},\ }\href@noop {} {\bibfield  {journal} {\bibinfo
  {journal} {arXiv preprint math/0303109}\ } (\bibinfo {year}
  {2003})}\BibitemShut {NoStop}%
\bibitem [{\citenamefont {Samal}\ \emph {et~al.}(2018)\citenamefont {Samal},
  \citenamefont {Sreejith}, \citenamefont {Gu}, \citenamefont {Liu},
  \citenamefont {Saucan},\ and\ \citenamefont {Jost}}]{samal2018comparative}%
  \BibitemOpen
  \bibfield  {author} {\bibinfo {author} {\bibfnamefont {A.}~\bibnamefont
  {Samal}}, \bibinfo {author} {\bibfnamefont {R.}~\bibnamefont {Sreejith}},
  \bibinfo {author} {\bibfnamefont {J.}~\bibnamefont {Gu}}, \bibinfo {author}
  {\bibfnamefont {S.}~\bibnamefont {Liu}}, \bibinfo {author} {\bibfnamefont
  {E.}~\bibnamefont {Saucan}},\ and\ \bibinfo {author} {\bibfnamefont
  {J.}~\bibnamefont {Jost}},\ }\bibfield  {title} {\bibinfo {title}
  {Comparative analysis of two discretizations of ricci curvature for complex
  networks},\ }\href@noop {} {\bibfield  {journal} {\bibinfo  {journal}
  {Scientific reports}\ }\textbf {\bibinfo {volume} {8}},\ \bibinfo {pages}
  {8650} (\bibinfo {year} {2018})}\BibitemShut {NoStop}%
\bibitem [{\citenamefont {Ollivier}(2007)}]{ollivier2007ricci}%
  \BibitemOpen
  \bibfield  {author} {\bibinfo {author} {\bibfnamefont {Y.}~\bibnamefont
  {Ollivier}},\ }\bibfield  {title} {\bibinfo {title} {Ricci curvature of
  metric spaces},\ }\href@noop {} {\bibfield  {journal} {\bibinfo  {journal}
  {Comptes Rendus Mathematique}\ }\textbf {\bibinfo {volume} {345}},\ \bibinfo
  {pages} {643} (\bibinfo {year} {2007})}\BibitemShut {NoStop}%
\bibitem [{\citenamefont {Ollivier}(2009)}]{ollivier2009ricci}%
  \BibitemOpen
  \bibfield  {author} {\bibinfo {author} {\bibfnamefont {Y.}~\bibnamefont
  {Ollivier}},\ }\bibfield  {title} {\bibinfo {title} {Ricci curvature of
  markov chains on metric spaces},\ }\href@noop {} {\bibfield  {journal}
  {\bibinfo  {journal} {Journal of Functional Analysis}\ }\textbf {\bibinfo
  {volume} {256}},\ \bibinfo {pages} {810} (\bibinfo {year}
  {2009})}\BibitemShut {NoStop}%
\bibitem [{\citenamefont {Forman}(2003)}]{forman2003bochner}%
  \BibitemOpen
  \bibfield  {author} {\bibinfo {author} {\bibnamefont {Forman}},\ }\bibfield
  {title} {\bibinfo {title} {Bochner's method for cell complexes and
  combinatorial ricci curvature},\ }\href@noop {} {\bibfield  {journal}
  {\bibinfo  {journal} {Discrete \& Computational Geometry}\ }\textbf {\bibinfo
  {volume} {29}},\ \bibinfo {pages} {323} (\bibinfo {year} {2003})}\BibitemShut
  {NoStop}%
\bibitem [{\citenamefont {Sreejith}\ \emph {et~al.}(2016)\citenamefont
  {Sreejith}, \citenamefont {Mohanraj}, \citenamefont {Jost}, \citenamefont
  {Saucan},\ and\ \citenamefont {Samal}}]{sreejith2016forman}%
  \BibitemOpen
  \bibfield  {author} {\bibinfo {author} {\bibfnamefont {R.}~\bibnamefont
  {Sreejith}}, \bibinfo {author} {\bibfnamefont {K.}~\bibnamefont {Mohanraj}},
  \bibinfo {author} {\bibfnamefont {J.}~\bibnamefont {Jost}}, \bibinfo {author}
  {\bibfnamefont {E.}~\bibnamefont {Saucan}},\ and\ \bibinfo {author}
  {\bibfnamefont {A.}~\bibnamefont {Samal}},\ }\bibfield  {title} {\bibinfo
  {title} {Forman curvature for complex networks},\ }\href@noop {} {\bibfield
  {journal} {\bibinfo  {journal} {Journal of Statistical Mechanics: Theory and
  Experiment}\ }\textbf {\bibinfo {volume} {2016}},\ \bibinfo {pages} {063206}
  (\bibinfo {year} {2016})}\BibitemShut {NoStop}%
\bibitem [{\citenamefont {Csaki}(1996)}]{csaki1996minimal}%
  \BibitemOpen
  \bibfield  {author} {\bibinfo {author} {\bibfnamefont {C.}~\bibnamefont
  {Csaki}},\ }\bibfield  {title} {\bibinfo {title} {The minimal supersymmetric
  standard model},\ }\href@noop {} {\bibfield  {journal} {\bibinfo  {journal}
  {Modern Physics Letters A}\ }\textbf {\bibinfo {volume} {11}},\ \bibinfo
  {pages} {599} (\bibinfo {year} {1996})}\BibitemShut {NoStop}%
\bibitem [{\citenamefont {Martin}(1998)}]{martin1998supersymmetry}%
  \BibitemOpen
  \bibfield  {author} {\bibinfo {author} {\bibfnamefont {S.~P.}\ \bibnamefont
  {Martin}},\ }\bibfield  {title} {\bibinfo {title} {A supersymmetry primer},\
  }in\ \href@noop {} {\emph {\bibinfo {booktitle} {Perspectives on
  supersymmetry}}}\ (\bibinfo  {publisher} {World Scientific},\ \bibinfo {year}
  {1998})\ pp.\ \bibinfo {pages} {1--98}\BibitemShut {NoStop}%
\bibitem [{\citenamefont {Baer}\ and\ \citenamefont
  {Tata}(2006)}]{Baer:2006rs}%
  \BibitemOpen
  \bibfield  {author} {\bibinfo {author} {\bibfnamefont {H.}~\bibnamefont
  {Baer}}\ and\ \bibinfo {author} {\bibfnamefont {X.}~\bibnamefont {Tata}},\
  }\href@noop {} {\emph {\bibinfo {title} {{Weak scale supersymmetry: From
  superfields to scattering events}}}}\ (\bibinfo  {publisher} {Cambridge
  University Press},\ \bibinfo {year} {2006})\BibitemShut {NoStop}%
\bibitem [{\citenamefont {Ellwanger}\ \emph {et~al.}(2010)\citenamefont
  {Ellwanger}, \citenamefont {Hugonie},\ and\ \citenamefont
  {Teixeira}}]{ellwanger2010next}%
  \BibitemOpen
  \bibfield  {author} {\bibinfo {author} {\bibfnamefont {U.}~\bibnamefont
  {Ellwanger}}, \bibinfo {author} {\bibfnamefont {C.}~\bibnamefont {Hugonie}},\
  and\ \bibinfo {author} {\bibfnamefont {A.~M.}\ \bibnamefont {Teixeira}},\
  }\bibfield  {title} {\bibinfo {title} {The next-to-minimal supersymmetric
  standard model},\ }\href@noop {} {\bibfield  {journal} {\bibinfo  {journal}
  {Physics Reports}\ }\textbf {\bibinfo {volume} {496}},\ \bibinfo {pages} {1}
  (\bibinfo {year} {2010})}\BibitemShut {NoStop}%
\bibitem [{\citenamefont {Cembranos}\ \emph {et~al.}(2007)\citenamefont
  {Cembranos}, \citenamefont {Feng},\ and\ \citenamefont
  {Strigari}}]{cembranos2007exotic}%
  \BibitemOpen
  \bibfield  {author} {\bibinfo {author} {\bibfnamefont {J.~A.}\ \bibnamefont
  {Cembranos}}, \bibinfo {author} {\bibfnamefont {J.~L.}\ \bibnamefont
  {Feng}},\ and\ \bibinfo {author} {\bibfnamefont {L.~E.}\ \bibnamefont
  {Strigari}},\ }\bibfield  {title} {\bibinfo {title} {Exotic collider signals
  from the complete phase diagram of minimal universal extra dimensions},\
  }\href@noop {} {\bibfield  {journal} {\bibinfo  {journal} {Physical Review
  D}\ }\textbf {\bibinfo {volume} {75}},\ \bibinfo {pages} {036004} (\bibinfo
  {year} {2007})}\BibitemShut {NoStop}%
\bibitem [{\citenamefont {Datta}\ \emph {et~al.}(2010)\citenamefont {Datta},
  \citenamefont {Kong},\ and\ \citenamefont {Matchev}}]{datta2010minimal}%
  \BibitemOpen
  \bibfield  {author} {\bibinfo {author} {\bibfnamefont {A.}~\bibnamefont
  {Datta}}, \bibinfo {author} {\bibfnamefont {K.}~\bibnamefont {Kong}},\ and\
  \bibinfo {author} {\bibfnamefont {K.~T.}\ \bibnamefont {Matchev}},\
  }\bibfield  {title} {\bibinfo {title} {Minimal universal extra dimensions in
  calchep/comphep},\ }\href@noop {} {\bibfield  {journal} {\bibinfo  {journal}
  {New Journal of Physics}\ }\textbf {\bibinfo {volume} {12}},\ \bibinfo
  {pages} {075017} (\bibinfo {year} {2010})}\BibitemShut {NoStop}%
\bibitem [{\citenamefont {Beuria}\ \emph {et~al.}(2018)\citenamefont {Beuria},
  \citenamefont {Datta}, \citenamefont {Debnath},\ and\ \citenamefont
  {Matchev}}]{beuria2018lhc}%
  \BibitemOpen
  \bibfield  {author} {\bibinfo {author} {\bibfnamefont {J.}~\bibnamefont
  {Beuria}}, \bibinfo {author} {\bibfnamefont {A.}~\bibnamefont {Datta}},
  \bibinfo {author} {\bibfnamefont {D.}~\bibnamefont {Debnath}},\ and\ \bibinfo
  {author} {\bibfnamefont {K.~T.}\ \bibnamefont {Matchev}},\ }\bibfield
  {title} {\bibinfo {title} {Lhc collider phenomenology of minimal universal
  extra dimensions},\ }\href@noop {} {\bibfield  {journal} {\bibinfo  {journal}
  {Computer Physics Communications}\ }\textbf {\bibinfo {volume} {226}},\
  \bibinfo {pages} {187} (\bibinfo {year} {2018})}\BibitemShut {NoStop}%
\bibitem [{\citenamefont {Davidson}\ and\ \citenamefont
  {Haber}(2005)}]{davidson2005basis}%
  \BibitemOpen
  \bibfield  {author} {\bibinfo {author} {\bibfnamefont {S.}~\bibnamefont
  {Davidson}}\ and\ \bibinfo {author} {\bibfnamefont {H.~E.}\ \bibnamefont
  {Haber}},\ }\bibfield  {title} {\bibinfo {title} {Basis-independent methods
  for the two-higgs-doublet model},\ }\href@noop {} {\bibfield  {journal}
  {\bibinfo  {journal} {Physical Review D}\ }\textbf {\bibinfo {volume} {72}},\
  \bibinfo {pages} {035004} (\bibinfo {year} {2005})}\BibitemShut {NoStop}%
\bibitem [{\citenamefont {Branco}\ \emph {et~al.}(2012)\citenamefont {Branco},
  \citenamefont {Ferreira}, \citenamefont {Lavoura}, \citenamefont {Rebelo},
  \citenamefont {Sher},\ and\ \citenamefont {Silva}}]{branco2012theory}%
  \BibitemOpen
  \bibfield  {author} {\bibinfo {author} {\bibfnamefont {G.~C.}\ \bibnamefont
  {Branco}}, \bibinfo {author} {\bibfnamefont {P.}~\bibnamefont {Ferreira}},
  \bibinfo {author} {\bibfnamefont {L.}~\bibnamefont {Lavoura}}, \bibinfo
  {author} {\bibfnamefont {M.}~\bibnamefont {Rebelo}}, \bibinfo {author}
  {\bibfnamefont {M.}~\bibnamefont {Sher}},\ and\ \bibinfo {author}
  {\bibfnamefont {J.~P.}\ \bibnamefont {Silva}},\ }\bibfield  {title} {\bibinfo
  {title} {Theory and phenomenology of two-higgs-doublet models},\ }\href@noop
  {} {\bibfield  {journal} {\bibinfo  {journal} {Physics reports}\ }\textbf
  {\bibinfo {volume} {516}},\ \bibinfo {pages} {1} (\bibinfo {year}
  {2012})}\BibitemShut {NoStop}%
\bibitem [{\citenamefont {Ham}\ \emph {et~al.}(2005)\citenamefont {Ham},
  \citenamefont {Jeong},\ and\ \citenamefont {Oh}}]{ham2005electroweak}%
  \BibitemOpen
  \bibfield  {author} {\bibinfo {author} {\bibfnamefont {S.}~\bibnamefont
  {Ham}}, \bibinfo {author} {\bibfnamefont {Y.}~\bibnamefont {Jeong}},\ and\
  \bibinfo {author} {\bibfnamefont {S.}~\bibnamefont {Oh}},\ }\bibfield
  {title} {\bibinfo {title} {Electroweak phase transition in an extension of
  the standard model with a real higgs singlet},\ }\href@noop {} {\bibfield
  {journal} {\bibinfo  {journal} {Journal of Physics G: Nuclear and Particle
  Physics}\ }\textbf {\bibinfo {volume} {31}},\ \bibinfo {pages} {857}
  (\bibinfo {year} {2005})}\BibitemShut {NoStop}%
\bibitem [{\citenamefont {Barger}\ \emph {et~al.}(2008)\citenamefont {Barger},
  \citenamefont {Langacker}, \citenamefont {McCaskey}, \citenamefont
  {Ramsey-Musolf},\ and\ \citenamefont {Shaughnessy}}]{barger2008cern}%
  \BibitemOpen
  \bibfield  {author} {\bibinfo {author} {\bibfnamefont {V.}~\bibnamefont
  {Barger}}, \bibinfo {author} {\bibfnamefont {P.}~\bibnamefont {Langacker}},
  \bibinfo {author} {\bibfnamefont {M.}~\bibnamefont {McCaskey}}, \bibinfo
  {author} {\bibfnamefont {M.~J.}\ \bibnamefont {Ramsey-Musolf}},\ and\
  \bibinfo {author} {\bibfnamefont {G.}~\bibnamefont {Shaughnessy}},\
  }\bibfield  {title} {\bibinfo {title} {Cern lhc phenomenology of an extended
  standard model with a real scalar singlet},\ }\href@noop {} {\bibfield
  {journal} {\bibinfo  {journal} {Physical Review D}\ }\textbf {\bibinfo
  {volume} {77}},\ \bibinfo {pages} {035005} (\bibinfo {year}
  {2008})}\BibitemShut {NoStop}%
\bibitem [{\citenamefont {Guo}\ and\ \citenamefont {Wu}(2010)}]{guo2010real}%
  \BibitemOpen
  \bibfield  {author} {\bibinfo {author} {\bibfnamefont {W.-L.}\ \bibnamefont
  {Guo}}\ and\ \bibinfo {author} {\bibfnamefont {Y.-L.}\ \bibnamefont {Wu}},\
  }\bibfield  {title} {\bibinfo {title} {The real singlet scalar dark matter
  model},\ }\href@noop {} {\bibfield  {journal} {\bibinfo  {journal} {Journal
  of High Energy Physics}\ }\textbf {\bibinfo {volume} {2010}},\ \bibinfo
  {pages} {1} (\bibinfo {year} {2010})}\BibitemShut {NoStop}%
\bibitem [{\citenamefont {Cohen-Steiner}\ \emph {et~al.}(2010)\citenamefont
  {Cohen-Steiner}, \citenamefont {Edelsbrunner}, \citenamefont {Harer},\ and\
  \citenamefont {Mileyko}}]{cohen2010lipschitz}%
  \BibitemOpen
  \bibfield  {author} {\bibinfo {author} {\bibfnamefont {D.}~\bibnamefont
  {Cohen-Steiner}}, \bibinfo {author} {\bibfnamefont {H.}~\bibnamefont
  {Edelsbrunner}}, \bibinfo {author} {\bibfnamefont {J.}~\bibnamefont
  {Harer}},\ and\ \bibinfo {author} {\bibfnamefont {Y.}~\bibnamefont
  {Mileyko}},\ }\bibfield  {title} {\bibinfo {title} {Lipschitz functions have
  l p-stable persistence},\ }\href@noop {} {\bibfield  {journal} {\bibinfo
  {journal} {Foundations of computational mathematics}\ }\textbf {\bibinfo
  {volume} {10}},\ \bibinfo {pages} {127} (\bibinfo {year} {2010})}\BibitemShut
  {NoStop}%
\bibitem [{\citenamefont {Skraba}\ and\ \citenamefont
  {Turner}(2020)}]{skraba2020wasserstein}%
  \BibitemOpen
  \bibfield  {author} {\bibinfo {author} {\bibfnamefont {P.}~\bibnamefont
  {Skraba}}\ and\ \bibinfo {author} {\bibfnamefont {K.}~\bibnamefont
  {Turner}},\ }\bibfield  {title} {\bibinfo {title} {Wasserstein stability for
  persistence diagrams},\ }\href@noop {} {\bibfield  {journal} {\bibinfo
  {journal} {arXiv preprint arXiv:2006.16824}\ } (\bibinfo {year}
  {2020})}\BibitemShut {NoStop}%
\bibitem [{\citenamefont {Songdechakraiwut}\ \emph {et~al.}(2023)\citenamefont
  {Songdechakraiwut}, \citenamefont {Krause}, \citenamefont {Banks},
  \citenamefont {Nourski},\ and\ \citenamefont
  {Van~Veen}}]{songdechakraiwut2023wasserstein}%
  \BibitemOpen
  \bibfield  {author} {\bibinfo {author} {\bibfnamefont {T.}~\bibnamefont
  {Songdechakraiwut}}, \bibinfo {author} {\bibfnamefont {B.~M.}\ \bibnamefont
  {Krause}}, \bibinfo {author} {\bibfnamefont {M.~I.}\ \bibnamefont {Banks}},
  \bibinfo {author} {\bibfnamefont {K.~V.}\ \bibnamefont {Nourski}},\ and\
  \bibinfo {author} {\bibfnamefont {B.~D.}\ \bibnamefont {Van~Veen}},\
  }\bibfield  {title} {\bibinfo {title} {Wasserstein distance-preserving vector
  space of persistent homology},\ }in\ \href@noop {} {\emph {\bibinfo
  {booktitle} {International Conference on Medical Image Computing and
  Computer-Assisted Intervention}}}\ (\bibinfo {organization} {Springer},\
  \bibinfo {year} {2023})\ pp.\ \bibinfo {pages} {277--286}\BibitemShut
  {NoStop}%
\bibitem [{\citenamefont {Das}\ \emph {et~al.}(2012)\citenamefont {Das},
  \citenamefont {Ellwanger},\ and\ \citenamefont {Teixeira}}]{das2012nmsdecay}%
  \BibitemOpen
  \bibfield  {author} {\bibinfo {author} {\bibfnamefont {D.}~\bibnamefont
  {Das}}, \bibinfo {author} {\bibfnamefont {U.}~\bibnamefont {Ellwanger}},\
  and\ \bibinfo {author} {\bibfnamefont {A.~M.}\ \bibnamefont {Teixeira}},\
  }\bibfield  {title} {\bibinfo {title} {Nmsdecay: a fortran code for
  supersymmetric particle decays in the next-to-minimal supersymmetric standard
  model},\ }\href@noop {} {\bibfield  {journal} {\bibinfo  {journal} {Computer
  Physics Communications}\ }\textbf {\bibinfo {volume} {183}},\ \bibinfo
  {pages} {774} (\bibinfo {year} {2012})}\BibitemShut {NoStop}%
\bibitem [{\citenamefont {Aad}\ \emph {et~al.}(2023)\citenamefont {Aad} \emph
  {et~al.}}]{ATLAS:2022zwa}%
  \BibitemOpen
  \bibfield  {author} {\bibinfo {author} {\bibfnamefont {G.}~\bibnamefont
  {Aad}} \emph {et~al.} (\bibinfo {collaboration} {ATLAS}),\ }\bibfield
  {title} {\bibinfo {title} {{ATLAS collaboration. "Searches for new phenomena
  in events with two leptons, jets, and missing transverse momentum in
  $139~\text {fb}^{-1} $ of $\sqrt {s}= 13~ $ TeV $ pp $ collisions with the
  ATLAS detector.}},\ }\href {https://doi.org/10.1140/epjc/s10052-023-11434-w}
  {\bibfield  {journal} {\bibinfo  {journal} {Eur. Phys. J. C}\ }\textbf
  {\bibinfo {volume} {83}},\ \bibinfo {pages} {515} (\bibinfo {year} {2023})},\
  \Eprint {https://arxiv.org/abs/2204.13072} {arXiv:2204.13072 [hep-ex]}
  \BibitemShut {NoStop}%
\bibitem [{\citenamefont {Alwall}\ \emph {et~al.}(2014)\citenamefont {Alwall},
  \citenamefont {Frederix}, \citenamefont {Frixione}, \citenamefont {Hirschi},
  \citenamefont {Maltoni}, \citenamefont {Mattelaer}, \citenamefont {Shao},
  \citenamefont {Stelzer}, \citenamefont {Torrielli},\ and\ \citenamefont
  {Zaro}}]{Alwall:2014hca}%
  \BibitemOpen
  \bibfield  {author} {\bibinfo {author} {\bibfnamefont {J.}~\bibnamefont
  {Alwall}}, \bibinfo {author} {\bibfnamefont {R.}~\bibnamefont {Frederix}},
  \bibinfo {author} {\bibfnamefont {S.}~\bibnamefont {Frixione}}, \bibinfo
  {author} {\bibfnamefont {V.}~\bibnamefont {Hirschi}}, \bibinfo {author}
  {\bibfnamefont {F.}~\bibnamefont {Maltoni}}, \bibinfo {author} {\bibfnamefont
  {O.}~\bibnamefont {Mattelaer}}, \bibinfo {author} {\bibfnamefont {H.~S.}\
  \bibnamefont {Shao}}, \bibinfo {author} {\bibfnamefont {T.}~\bibnamefont
  {Stelzer}}, \bibinfo {author} {\bibfnamefont {P.}~\bibnamefont {Torrielli}},\
  and\ \bibinfo {author} {\bibfnamefont {M.}~\bibnamefont {Zaro}},\ }\bibfield
  {title} {\bibinfo {title} {{The automated computation of tree-level and
  next-to-leading order differential cross sections, and their matching to
  parton shower simulations}},\ }\href
  {https://doi.org/10.1007/JHEP07(2014)079} {\bibfield  {journal} {\bibinfo
  {journal} {JHEP}\ }\textbf {\bibinfo {volume} {07}},\ \bibinfo {pages}
  {079}},\ \Eprint {https://arxiv.org/abs/1405.0301} {arXiv:1405.0301 [hep-ph]}
  \BibitemShut {NoStop}%
\bibitem [{\citenamefont {Frederix}\ \emph {et~al.}(2018)\citenamefont
  {Frederix}, \citenamefont {Frixione}, \citenamefont {Hirschi}, \citenamefont
  {Pagani}, \citenamefont {Shao},\ and\ \citenamefont
  {Zaro}}]{Frederix:2018nkq}%
  \BibitemOpen
  \bibfield  {author} {\bibinfo {author} {\bibfnamefont {R.}~\bibnamefont
  {Frederix}}, \bibinfo {author} {\bibfnamefont {S.}~\bibnamefont {Frixione}},
  \bibinfo {author} {\bibfnamefont {V.}~\bibnamefont {Hirschi}}, \bibinfo
  {author} {\bibfnamefont {D.}~\bibnamefont {Pagani}}, \bibinfo {author}
  {\bibfnamefont {H.~S.}\ \bibnamefont {Shao}},\ and\ \bibinfo {author}
  {\bibfnamefont {M.}~\bibnamefont {Zaro}},\ }\bibfield  {title} {\bibinfo
  {title} {{The automation of next-to-leading order electroweak
  calculations}},\ }\href {https://doi.org/10.1007/JHEP11(2021)085} {\bibfield
  {journal} {\bibinfo  {journal} {JHEP}\ }\textbf {\bibinfo {volume} {07}},\
  \bibinfo {pages} {185}},\ \bibinfo {note} {[Erratum: JHEP 11, 085 (2021)]},\
  \Eprint {https://arxiv.org/abs/1804.10017} {arXiv:1804.10017 [hep-ph]}
  \BibitemShut {NoStop}%
\bibitem [{\citenamefont {Ball}\ \emph {et~al.}(2017)\citenamefont {Ball},
  \citenamefont {Bertone}, \citenamefont {Carrazza}, \citenamefont {Debbio},
  \citenamefont {Forte}, \citenamefont {Groth-Merrild}, \citenamefont
  {Guffanti}, \citenamefont {Hartland}, \citenamefont {Kassabov}, \citenamefont
  {Latorre} \emph {et~al.}}]{ball2017parton}%
  \BibitemOpen
  \bibfield  {author} {\bibinfo {author} {\bibfnamefont {R.~D.}\ \bibnamefont
  {Ball}}, \bibinfo {author} {\bibfnamefont {V.}~\bibnamefont {Bertone}},
  \bibinfo {author} {\bibfnamefont {S.}~\bibnamefont {Carrazza}}, \bibinfo
  {author} {\bibfnamefont {L.~D.}\ \bibnamefont {Debbio}}, \bibinfo {author}
  {\bibfnamefont {S.}~\bibnamefont {Forte}}, \bibinfo {author} {\bibfnamefont
  {P.}~\bibnamefont {Groth-Merrild}}, \bibinfo {author} {\bibfnamefont
  {A.}~\bibnamefont {Guffanti}}, \bibinfo {author} {\bibfnamefont {N.~P.}\
  \bibnamefont {Hartland}}, \bibinfo {author} {\bibfnamefont {Z.}~\bibnamefont
  {Kassabov}}, \bibinfo {author} {\bibfnamefont {J.~I.}\ \bibnamefont
  {Latorre}}, \emph {et~al.},\ }\bibfield  {title} {\bibinfo {title} {Parton
  distributions from high-precision collider data: Nnpdf collaboration},\
  }\href@noop {} {\bibfield  {journal} {\bibinfo  {journal} {The European
  Physical Journal C}\ }\textbf {\bibinfo {volume} {77}},\ \bibinfo {pages} {1}
  (\bibinfo {year} {2017})}\BibitemShut {NoStop}%
\bibitem [{\citenamefont {Bierlich}\ \emph {et~al.}(2022)\citenamefont
  {Bierlich}, \citenamefont {Chakraborty}, \citenamefont {Desai}, \citenamefont
  {Gellersen}, \citenamefont {Helenius}, \citenamefont {Ilten}, \citenamefont
  {L{\"o}nnblad}, \citenamefont {Mrenna}, \citenamefont {Prestel},
  \citenamefont {Preuss} \emph {et~al.}}]{bierlich2022comprehensive}%
  \BibitemOpen
  \bibfield  {author} {\bibinfo {author} {\bibfnamefont {C.}~\bibnamefont
  {Bierlich}}, \bibinfo {author} {\bibfnamefont {S.}~\bibnamefont
  {Chakraborty}}, \bibinfo {author} {\bibfnamefont {N.}~\bibnamefont {Desai}},
  \bibinfo {author} {\bibfnamefont {L.}~\bibnamefont {Gellersen}}, \bibinfo
  {author} {\bibfnamefont {I.}~\bibnamefont {Helenius}}, \bibinfo {author}
  {\bibfnamefont {P.}~\bibnamefont {Ilten}}, \bibinfo {author} {\bibfnamefont
  {L.}~\bibnamefont {L{\"o}nnblad}}, \bibinfo {author} {\bibfnamefont
  {S.}~\bibnamefont {Mrenna}}, \bibinfo {author} {\bibfnamefont
  {S.}~\bibnamefont {Prestel}}, \bibinfo {author} {\bibfnamefont {C.~T.}\
  \bibnamefont {Preuss}}, \emph {et~al.},\ }\bibfield  {title} {\bibinfo
  {title} {A comprehensive guide to the physics and usage of pythia 8.3},\
  }\href@noop {} {\bibfield  {journal} {\bibinfo  {journal} {SciPost Physics
  Codebases}\ ,\ \bibinfo {pages} {008}} (\bibinfo {year} {2022})}\BibitemShut
  {NoStop}%
\bibitem [{\citenamefont {Cacciari}\ and\ \citenamefont
  {Salam}(2006)}]{Cacciari:2005hq}%
  \BibitemOpen
  \bibfield  {author} {\bibinfo {author} {\bibfnamefont {M.}~\bibnamefont
  {Cacciari}}\ and\ \bibinfo {author} {\bibfnamefont {G.~P.}\ \bibnamefont
  {Salam}},\ }\bibfield  {title} {\bibinfo {title} {{Dispelling the $N^{3}$
  myth for the $k_t$ jet-finder}},\ }\href
  {https://doi.org/10.1016/j.physletb.2006.08.037} {\bibfield  {journal}
  {\bibinfo  {journal} {Phys. Lett. B}\ }\textbf {\bibinfo {volume} {641}},\
  \bibinfo {pages} {57} (\bibinfo {year} {2006})},\ \Eprint
  {https://arxiv.org/abs/hep-ph/0512210} {arXiv:hep-ph/0512210} \BibitemShut
  {NoStop}%
\bibitem [{\citenamefont {Cacciari}\ \emph {et~al.}(2012)\citenamefont
  {Cacciari}, \citenamefont {Salam},\ and\ \citenamefont
  {Soyez}}]{Cacciari:2011ma}%
  \BibitemOpen
  \bibfield  {author} {\bibinfo {author} {\bibfnamefont {M.}~\bibnamefont
  {Cacciari}}, \bibinfo {author} {\bibfnamefont {G.~P.}\ \bibnamefont
  {Salam}},\ and\ \bibinfo {author} {\bibfnamefont {G.}~\bibnamefont {Soyez}},\
  }\bibfield  {title} {\bibinfo {title} {{FastJet User Manual}},\ }\href
  {https://doi.org/10.1140/epjc/s10052-012-1896-2} {\bibfield  {journal}
  {\bibinfo  {journal} {Eur. Phys. J. C}\ }\textbf {\bibinfo {volume} {72}},\
  \bibinfo {pages} {1896} (\bibinfo {year} {2012})},\ \Eprint
  {https://arxiv.org/abs/1111.6097} {arXiv:1111.6097 [hep-ph]} \BibitemShut
  {NoStop}%
\bibitem [{\citenamefont {de~Favereau}\ \emph {et~al.}(2014)\citenamefont
  {de~Favereau}, \citenamefont {Delaere}, \citenamefont {Demin}, \citenamefont
  {Giammanco}, \citenamefont {Lema\^\i{}tre}, \citenamefont {Mertens},\ and\
  \citenamefont {Selvaggi}}]{deFavereau:2013fsa}%
  \BibitemOpen
  \bibfield  {author} {\bibinfo {author} {\bibfnamefont {J.}~\bibnamefont
  {de~Favereau}}, \bibinfo {author} {\bibfnamefont {C.}~\bibnamefont
  {Delaere}}, \bibinfo {author} {\bibfnamefont {P.}~\bibnamefont {Demin}},
  \bibinfo {author} {\bibfnamefont {A.}~\bibnamefont {Giammanco}}, \bibinfo
  {author} {\bibfnamefont {V.}~\bibnamefont {Lema\^\i{}tre}}, \bibinfo {author}
  {\bibfnamefont {A.}~\bibnamefont {Mertens}},\ and\ \bibinfo {author}
  {\bibfnamefont {M.}~\bibnamefont {Selvaggi}} (\bibinfo {collaboration}
  {DELPHES 3}),\ }\bibfield  {title} {\bibinfo {title} {{DELPHES 3, A modular
  framework for fast simulation of a generic collider experiment}},\ }\href
  {https://doi.org/10.1007/JHEP02(2014)057} {\bibfield  {journal} {\bibinfo
  {journal} {JHEP}\ }\textbf {\bibinfo {volume} {02}},\ \bibinfo {pages}
  {057}},\ \Eprint {https://arxiv.org/abs/1307.6346} {arXiv:1307.6346 [hep-ex]}
  \BibitemShut {NoStop}%
\bibitem [{\citenamefont {Guo}\ \emph {et~al.}(2024)\citenamefont {Guo},
  \citenamefont {Feng}, \citenamefont {Di}, \citenamefont {Lu},\ and\
  \citenamefont {Yang}}]{guo2024mlanalysis}%
  \BibitemOpen
  \bibfield  {author} {\bibinfo {author} {\bibfnamefont {Y.-C.}\ \bibnamefont
  {Guo}}, \bibinfo {author} {\bibfnamefont {F.}~\bibnamefont {Feng}}, \bibinfo
  {author} {\bibfnamefont {A.}~\bibnamefont {Di}}, \bibinfo {author}
  {\bibfnamefont {S.-Q.}\ \bibnamefont {Lu}},\ and\ \bibinfo {author}
  {\bibfnamefont {J.-C.}\ \bibnamefont {Yang}},\ }\bibfield  {title} {\bibinfo
  {title} {Mlanalysis: An open-source program for high energy physics
  analyses},\ }\href@noop {} {\bibfield  {journal} {\bibinfo  {journal}
  {Computer Physics Communications}\ }\textbf {\bibinfo {volume} {294}},\
  \bibinfo {pages} {108957} (\bibinfo {year} {2024})}\BibitemShut {NoStop}%
\bibitem [{\citenamefont {Tauzin}\ \emph {et~al.}(2021)\citenamefont {Tauzin},
  \citenamefont {Lupo}, \citenamefont {Tunstall}, \citenamefont {P\'{e}rez},
  \citenamefont {Caorsi}, \citenamefont {Medina-Mardones}, \citenamefont
  {Dassatti},\ and\ \citenamefont {Hess}}]{giotto-tda}%
  \BibitemOpen
  \bibfield  {author} {\bibinfo {author} {\bibfnamefont {G.}~\bibnamefont
  {Tauzin}}, \bibinfo {author} {\bibfnamefont {U.}~\bibnamefont {Lupo}},
  \bibinfo {author} {\bibfnamefont {L.}~\bibnamefont {Tunstall}}, \bibinfo
  {author} {\bibfnamefont {J.~B.}\ \bibnamefont {P\'{e}rez}}, \bibinfo {author}
  {\bibfnamefont {M.}~\bibnamefont {Caorsi}}, \bibinfo {author} {\bibfnamefont
  {A.~M.}\ \bibnamefont {Medina-Mardones}}, \bibinfo {author} {\bibfnamefont
  {A.}~\bibnamefont {Dassatti}},\ and\ \bibinfo {author} {\bibfnamefont
  {K.}~\bibnamefont {Hess}},\ }\bibfield  {title} {\bibinfo {title}
  {giotto-tda: A topological data analysis toolkit for machine learning and
  data exploration},\ }\href {http://jmlr.org/papers/v22/20-325.html}
  {\bibfield  {journal} {\bibinfo  {journal} {Journal of Machine Learning
  Research}\ }\textbf {\bibinfo {volume} {22}},\ \bibinfo {pages} {1} (\bibinfo
  {year} {2021})}\BibitemShut {NoStop}%
\end{thebibliography}%

\end{document}